\documentclass[sn-mathphys-num]{sn-jnl}% Math and Physical Sciences Numbered Reference Style 

\usepackage{graphicx}%
\usepackage{multirow}%
\usepackage{amsmath,amssymb,amsfonts}%
\usepackage{amsthm}%
\usepackage{mathrsfs}%
\usepackage[title]{appendix}%
\usepackage{xcolor}%
\usepackage{textcomp}%
\usepackage{manyfoot}%
\usepackage{booktabs}%
\usepackage{algorithm}%
\usepackage{algorithmicx}%
\usepackage{algpseudocode}%
\usepackage{listings}%
\theoremstyle{thmstyleone}%
\theoremstyle{thmstyletwo}%

\theoremstyle{thmstylethree}%

\begin{document}

\title [mode = title]{Decoherence of the pair entangled states at thermalization and interparticle interaction. \\
	Heisenberg--Wigner path integral Monte Carlo simulations}  

\author*[1]{\fnm{Vladimir} \sur{Filinov}}\email{vladimir\_filinov@mail.ru}
\author[1]{\fnm{Pavel} \sur{ Levashov}}\email{pavelrlevashov@gmail.com}
\equalcont{These authors contributed equally to this work.}
\author[1]{\fnm{Alexander} \sur{Larkin}}\email{alexanderlarkin@rambler.ru}
\equalcont{These authors contributed equally to this work.}
\affil*[1]{\orgname{Joint Institute for High Temperatures of the Russian Academy of Sciences}, \orgaddress{\street{Izhorskaya 13, Bldg 2}, \city{Moscow}, \postcode{127412}, \country{Russia}}}

\abstract{
For	deeper understanding of quantum non-local correlations  
a new path integral \textit{real--time} quantum dynamics approach has been developed and applied to the  Monte Carlo simulation 
of the decoherence of pair entangled states in many--body systems in the framework of the ``replica trick''. 
Calculations have been carried out for the Heisenberg--Wigner path integral representation of the entangled state correlation function 
(ESCF) in the modified unified electron gas model as example. The Wiener--Khinchin transformation of ESCF has been applied to random paths 
in path integral allowing for the Monte Carlo simulation of ESCF. Presented results include the spin--resolved radial distribution 
functions, ESCFs and estimations of the decoherence time of the entangled electron pairs 
due to  thermalization and interparticle interaction.  
}




\maketitle

\section*{Introduction}%\label{sec1}

Entanglement represents the correlation property of quantum states of many-body systems and is important for understanding 
of  non-local quantum correlations 
\cite{sanctuary2006correlations,sanctuary2003interpretation}. 
There has been increasing interest to quantum systems described  by the density matrix $e^{-\beta\hat{H}}$ 
\cite{hornberger2009introduction,humeniuk2012quantum,melko2010finite}  
($\beta=1/T$ is inverse temperature and $\hat{H}$ is the Hamiltonian of the system)
and calculations of Renyi entanglement entropy (RE), which should be a maximum for a maximally entangled states
of a quantum many-body system. 
These calculations imply the necessity to reconstruct the reduced density matrix of a 
many--body systems using quantum Monte Carlo technique in the framework of ``replica trick'' 
\cite{hornberger2009introduction,calabrese2004entanglement,melko2010finite,humeniuk2012quantum,buividovich2008numerical}. 
The basic idea is to perform the quantum Monte Carlo simulations (QMC) with the action of the
\textit{imaginary-time}  propagator $e^{-\beta\hat{H}}$ in an extended ensemble for several 
replicas of the system and to redefine the simulation QMC box into periodical topology 
\cite{hornberger2009introduction,calabrese2004entanglement,melko2010finite,humeniuk2012quantum,buividovich2008numerical}. 

The present study is focused on the \textit{real--time} quantum simulations of decoherence decay of pair entangled states 
in a modified uniform electron gas model (UEG) being a two-component Coulomb system 
of quantum entangled electron pairs and 
uncorrelated positive charges simulating a neutralizing background. 

According to the basic ideas of the ``replica trick'' in path integral approach \cite{calabrese2004entanglement}   
extensively used both in the field theory analytically \cite{melko2010finite}  
and in Monte Carlo simulations of lattice gauge theories  \cite{buividovich2008numerical} 
we consider pairs of entangled electrons as particles connected by the paths generated by the imaginary-time propagator  
$e^{-\beta\hat{H}}$. 
Then for the quantum non-local density-density correlation function we  
use the analytical continuation of this propagator in the form 
$ e^{ {\rm i}\hat{H}t_{\rm c}} \mbox{ at }   t_{\rm c}=  t + {\rm i} \beta $ 
resulting to the Heisenberg representation of the entangled state correlation function. 
Panel a) in Fig.~1 shows several  pairs of entangled electrons by the paths generated by 
Heisenberg propagators 
$ e^{ {\rm i}\hat{H}t_{\rm c}} \mbox{ at }   t_{\rm c}=  t + {\rm i} \beta $, 
which includes also density matrix. 

Decoherence of the entangled pairs due to thermalization and interparticle interaction is described by 
the quantum non-local density--density correlation function  
represented by a developed Heisenberg--Wigner path integrals using analytical continuation of propagator 
$ e^{ {\rm i}\hat{H}t_{\rm c}} \mbox{ at }   t_{\rm c}=  t + {\rm i} \beta $ 
instead of the standard one  in literature $e^{-\beta\hat{H}}$ \cite{hornberger2009introduction,melko2010finite,humeniuk2012quantum}).  
We perform the quantum Monte Carlo (QMC) simulation of the entangled electron pairs in a ``replica''  extended ensemble  
\cite{hornberger2009introduction,melko2010finite,humeniuk2012quantum,calabrese2004entanglement,buividovich2008numerical}. 

From the physical point of view the density--density correlation function is a function describing scattering by unperturbed system \cite{sturm1993dynamic} 
as well as related correlations and dynamic characteristics, which may be important for experimental diagnostics.  

Our studies are based on a new \textit{real--time} dynamics approach that significantly differs from 
existing ones in the literature. 
The \textit{real--time} dynamics Feynman path integral Monte Carlo approach derived before \cite{alexandru2016monte} 
is restricted by the Feynman integrals sign problem 
that emerges from a highly oscillatory integrand of the path integral. The key idea of \cite{alexandru2016monte} is to deform 
the path integration domain to a complex manifold where the phase oscillations are mild and the \textit{sign problem} is manageable. 

Analogously in Monte Carlo studies \cite{filinov2020lefschetz}  the basic idea is to choose the Lefschetz thimbles 
as manifolds approaching the saddle point of the integrand in Feynman integrals. On thimble the imaginary part of 
the complex-valued action remains constant. 
As a result the integrand on each thimble does not oscillate, so the the \textit{sign problem} disappears and the integral can be 
calculated much more effectively. This approach allows also the Monte Carlo finding saddle points in 
the complexified domain of path integral integration. However this approaches requires analytical continuation of 
real  effective interparticle potential or pseudopotential on the complex manifold, what is often  non unique due to 
the practically empiric nature of interparticle interaction. 

To avoid complex-valued interparticle potentials in analytical continuation 
a generalization of the filtering procedure \cite{filinov1986calculation} was proposed for the practical solution of the 
Feynman integrals \textit{sign problem}, in which explicit oscillations effectively cancel in the oscillatory part of the original 
integrand \cite{wang2001generalized}. 

An advantage of the developed here approach is that it describes quantum real–time dynamics of a system and its related dynamic characteristics 
(such as, for example,  density–density correlation function) by real-valued functions and is free from 
the Feynman complex-valued path integral sign problem. This approach combines the Heisenberg and Wigner formulations of quantum mechanics, 
path integral technique and Wiener--Khinchin theorem 
applied to spectral densities of the paths in path integrals. As a result, the density--density correlation function expressed through 
the Weyl symbols of the density operators and the generalized Wigner function is represented as a path integral in phase space. 
As a consequence the developed method offers a wide possibility in treatment dynamic properties as it is formulated in the phase space and explicitly uses the momenta and coordinates of particles. 

Our approach differs significantly from the most traditional Wigner techniques reviewed in 
\cite{Tatr,gaspard1995new,langouche2013functional,jain2004semiclassical}, where the general background of the Weyl--Wigner representation of 
quantum mechanics, semiclassical methods, path integral representation of propagators, $\hbar$-expansion and the relation 
between quantum and classical mechanics have been discussed. Let us note that traditional techniques often  presuppose 
the existence of small physical parameters, so quantum statistics and wave interference effects can hardly describe  
subtle peculiarities of the most interesting strongly coupled quantum systems. 

The main difficulties of the existing path integral Monte Carlo approch (PIMC) for simulations of Bose and Fermi systems are 
necessities of symmetrization or antisymmetrization of the density operator, which 
implies the calculation of the sum over all particle permutations. A perspective approach 
to reduce this problem for bosons was developed in \cite{hirshberg2019path} by using path integral molecular
dynamics, which  allows enumerating all ring permutation configurations arising due to 
permutations of identical particles. In this approach the potential
and forces at each imaginary time step can be evaluated by using a recurrence relation which avoids 
enumerating all permutations, while providing the correct thermal expectation values.

On the other hand the main disadvantage of the existing PIMC for simulating Fermi systems is the notorious ``fermionic sign problem'' 
arising due to the antisymmetrization of a fermion density matrix \cite{EbelForFil,ceperley1991fermion,ceperley1992path}. 
To overcome this issue a lot of approaches were developed (see, for example,  
\cite{zamalin1977monte,EbelForFil,ForFilLarEbl,runeson2018quantum,dornheim2020attenuating}).  
In \cite{ceperley1991fermion,ceperley1992path} to avoid the ``fermionic sign problem'', a restricted fixed--node path--integral 
Monte Carlo (RPIMC) approach was offered. 
In RPIMC only positive permutations are taken into account, so the accuracy of the results is unknown.  

To correctly solve the ``fermionic sign problem''  
the exchange interaction of fermions can be presented in the form of a positive semi-definite 
Gram determinant \cite{filinov2022solution}. The verification of this approach was carried out by comparison of the calculated equation of state with experimental data for quantum and degenerate $^3 {\rm He}$. 

An alternative approach based on the Wigner formulation of quantum mechanics in the phase space 
\cite{wigner1934interaction,Tatr} was used in Refs.~\cite{larkin2017pauli,larkin2017peculiarities} to 
avoid the antisymmetrization of matrix elements and hence the ``sign problem''. This approach allows  
to reproduce the Pauli blocking of fermions and is able to calculate quantum momentum distribution functions
as well as transport properties \cite{EbelForFil}. 
However, the approach is not applicable at high degeneracy. 

In this paper we continue developing  the Wigner path integral Monte Carlo (WPIMC) technique 
\cite{larkin2017pauli,larkin2017peculiarities,filinov2025wigner,filinov2023density} applying it 
to the quantum simulations of decoherence decay of pair entangled states in modified UEG model. 

The paper is organized as follows.  

In section 1 we introduced  a modified UEG model being a two-component Coulomb system. 
of quantum entangled electron pairs and uncorrelated positive charges simulating a neutralizing background. 

In section 2 we developed the new Heisenberg--Wigner path integral representation of the 
the entangled state correlation function (ESCF)  
in the framework of the ``replica trick'' approach 
\cite{hornberger2009introduction,melko2010finite,humeniuk2012quantum,calabrese2004entanglement}. 

In section 3 we discuss  Heisenberg--Wigner path integral representation and  ``replica trick''. 

In section 4 we generalize  the Wiener--Khinchin theorem  allowing for Monte Carlo calculations of ESCF. 

%In section 4 we present the Wiener--Khinchin transformation of path integrals allowing for Monte Carlo calculations of ESCF. 

In section 5 we discuss results of our calculations presenting the spin--resolved radial distribution functions, 
ESCFs, the time entangled state correlation functions (TESCFs) 
and discuss decoherence time of the entangled pairs of electron due to  thermalization and interparticle interaction. 

In section 6 we summarize the basic ideas of developed approach and its possible perspectives.  
Details of carried out analytical transformations and Monte Carli simulation are discussed in Supplemental materials.  

\section{Entangled pairs and density structure factor  }\label{hf}  
Here for consideration of quantum  of non-local correlations  we are following to the basic ideas of the 
``replica trick'' \cite{calabrese2004entanglement} 
which has been  extensively used in field-theory both analytically \cite{metlitski2009entanglement} and as well as 
in Monte Carlo simulations of lattice gauge theories \cite{buividovich2008numerical}. 
We are using analytical continuation of propagator 
$ e^{ {\rm i}\hat{H}t_{\rm c}} \mbox{ at }   t_{\rm c}=  t + {\rm i} \beta $ 
instead of the standard one  in literature $e^{-\beta\hat{H}}$.  

Let the considered two-component Coulomb system in canonical ensemble is described by the Hamiltonian 
$\hat H=\hat K+\hat U$ ($\hat K$ and $\hat U$ are kinetic and potential energies operators).   
To analyze the spatial and dynamic characteristics of the pair entangled states decoherence let us consider 
the ESCF in Heisenberg representation (like diagonal elements of the density--density structure factor 
\cite{calabrese2004entanglement,sturm1993dynamic}) 

The ESCF in Heisenberg representation is presented in the form of the space and time Fourier transform of trace of density 
operators 
and includes averaging over configurations in the canonical ensemble taken as initial condition for 
infinite in time dynamics of virtual trajectories described by the propagator $e^{ {\rm i}\hat{H}t_{\rm c}}$
\begin{eqnarray}	
	%\begin{equation}
	S(k, \omega)=  
	Z^{-2}  \frac{1}{N}\sum_l 
	\int_{-\infty}^{\infty}  {\rm d} t \,  e^{-{\rm i} \omega  t } 
%	\nonumber\\	\times 
    \mbox{Tr}\left(  e^{-i\langle k |\underline{ \hat Q}_l } e^{ {\rm i}\hat{H}t_{\rm c}} e^{i\langle k | \hat Q_l\rangle}  
	e^{{-\rm i}\hat{H}t_{\rm c}^{*}} \right)   \, 
	%\nonumber\\&&
	\label{CFA1}
\end{eqnarray}		
%\end{equation}
%\end{multline}
%\end{widetext}
$\beta =1/T$ is the reciprocal temperature, $t_{\rm c}=  t + {\rm i} \beta $ is  a complex-valued ``time'', 
$\rm i$ is the imaginary unit, sum over $l$ is the sum over entangled pairs of  electrons from $1$ up to $N$, 
$\underline{\hat Q_l}$ and $\hat Q_l$  are  vector operators of coordinates of  two  entangled electrons in  pair number $l$  
\cite{melko2010finite,humeniuk2012quantum,calabrese2004entanglement} corresponding to ``inital time'' 
at $t=0$. The factors $e^{-i\langle k|\hat Q_l\rangle}$ in the sum 
$\frac{1}{\sqrt{N}}\sum_l e^{-i\langle k|\underline{ \hat Q}_l \rangle} =\hat\rho(k, \underline{\hat Q)} $ 
(and $\frac{1}{\sqrt{N}}\sum_l e^{-i\langle k|\hat Q_l\rangle}=\hat\rho(k, \hat Q) $ ) are defined by 
the Fourier transforms of the density operators $\hat{\rho}(\hat Q) = \frac{1}{\sqrt{N}}\sum_l \delta(\hat Q - \hat Q_l)$ 
at ``inital time'' ($t=0$). 

On the other hand, 
the term  $e^{ {\rm i}\hat{H}t_{\rm c}} e^{i \langle k|  \hat Q_l \rangle} e^{{-\rm i}\hat{H}t_{\rm c}^{*}} $ 
in Heisenberg representation corresponds to ``time'' $t_{\rm c}$, $k$ is a $3D$ vector.  
Here due to the ``replica trick'' \cite{calabrese2004entanglement} the normalization is $Z^{-2}$ 
\cite{calabrese2004entanglement,melko2010finite}, where 
$Z\left( N,V,T \right) =\mbox{Tr}\big(\hat{\rho}_0 \big)=\mbox{Tr}\big(e^{-\beta\hat{H}}\big)$ is the canonical partition function, 

We are going to apply the basic concepts of the Wigner formulation of quantum mechanics and path integral technique 
to  $S(k,\omega)$. The Wigner representation of  $S(k, \omega)$ 
can be identically rewritten in the form that includes the Weyl symbols of the density operator $\hat{\rho}$ and 
Fourier transform $W\left(\overline{PQ};\widetilde{PQ} ;\omega \right)$ of the 
antisymmetrized generalization of the Wigner function (see for details Supplemental material) 
\begin{multline}
	S(k, \omega) = Z^{-2} \frac{1}{N}\sum_l  \hspace{-4pt}\int {\rm d} t \,e^{-{\rm i} \omega t}\, 
	\hspace{-4pt}\int{\rm d}\overline{q} \, {\rm d} \overline{\overline{q}} \, {\rm d}\widetilde{q} \,{\rm d} \tilde{\tilde{q}} \, %\times {}\\
	\left\langle \overline{q}\left |e^{-i\langle k|\underline{ \hat Q}_l \rangle}  \right| \overline{\overline{q}} \right\rangle  
	\left\langle  \overline{\overline{q}}  \left|e^{{\rm i}\hat{H}t_{\rm c}}\right|
	\widetilde{q} \right\rangle \times {}\\
	\left\langle \widetilde{q} \left |  e^{i\langle k|\hat Q_l\rangle} 
	\right| \tilde{\tilde{q}} \right\rangle 
	\left\langle  \tilde{\tilde{q}} \left|e^{-{\rm i} \hat{H}t^*_{\rm c}}\right| \overline{q}  \right\rangle
	=(2\pi)^{-12N} \times {} %\\  
	\frac{1}{N}\sum_l  
	\hspace{-4pt} \int\hspace{-3pt} {\rm d}\overline{PQ} \, {\rm d}\widetilde{PQ}
	\,e^{i\langle k|\overline{Q}_l\rangle} e^{-i\langle k|\widetilde{Q}_l \rangle}  \, 
	W\left(\overline{PQ};\widetilde{PQ} ;\omega \right)
	%	\nonumber
	\label{CFA2}
\end{multline}
where $W\left(\overline{PQ};\widetilde{PQ} ;\omega \right)  =Z^{-2} \int {\rm d} t \,e^{-{\rm i} \omega t}\,  
\int\int {\rm d}\overline{\xi}\, {\rm d}\widetilde{\xi}\, 
e^{{\rm i} \langle \overline{P}| \overline{\xi}\rangle}  e^{{\rm i}\langle\widetilde{P}|\widetilde{\xi}\rangle} 
\left\langle \overline{Q}+\frac{\overline{\xi}}{2} 
\left|e^{{\rm i}\hat{H}t_{\rm c}}\right|
\widetilde{Q}-\frac{\widetilde{\xi}}{2}  \right\rangle 	\\ \times 
\left\langle \widetilde{Q}+\frac{\widetilde{\xi}}{2} \left|e^{-{\rm i} \hat{H}t^*_{\rm c}}\right|
\overline{Q}-\frac{\overline{\xi}}{2}  \right\rangle  
$
and we introduced the new variables $\overline{Q}=(\overline{q}+\overline{\overline{q}})/2$,  
$\widetilde{Q} =(\tilde{q}+\tilde{\tilde{q}} )/2$, 
$\overline\xi = \overline{\overline{q}} - \overline{q}$,   
$\tilde{\xi} = \tilde{\tilde{q}} - \tilde{q}$ 
and a short-hand notation  for 6$N$-dimensional phase space points 
$\overline{P}$, $\overline{Q}$ and $\widetilde{P}$, $\widetilde{Q}$ being  
momenta and coordinates of $2N$ electrons in the UEG model. 

We use the standard Dirac notations: operators are indicated with the sign ``hat'' for coordinate and momentum operators $\hat{q}$  and $\hat{p}$ . 
Here $|q \rangle $  are $3N$-dimensional vectors, which are eigenvectors of the position operator  
$\hat{q} |q\rangle =q |q\rangle$  \cite{Tatr},  
satisfying the conditions $\langle \tilde{q} | q\rangle=\delta(\tilde{q}-q)$, $\hat{I}=  \int {\rm d} q \, | q\rangle| \langle q |$ , 
$\hat{I}$ is the unit operator; $ |p\rangle$ are analogous eigenvectors of the operator $\hat{p}$, $\hat{p} |p \rangle =p |p\rangle$
$\langle \tilde{p} | p\rangle=\delta(\tilde{p}-p)$, $\hat{I}=| p \rangle \int {\rm d} p   \langle p |$. 
The commutator $ [ \hat{q} , \hat{p} ] = {\rm i}$ (for units with $\hbar=1$), 
$\langle q | p\rangle= (2 \pi)^{-N/2} \exp {\rm i}\langle q | p \rangle $. 
The angular brackets in the expression $\langle q_1|\hat{A}|q\rangle $ mean the scalar product of vectors 
$|q_1\rangle$ and  $|\hat{A}|q\rangle$. 
Here we imply that energy is expressed in units of $k_B T$ ($T$ is the temperature of the system. 

As $e^{i\langle k| \widetilde Q_l\rangle} $ and $e^{i\langle k|\overline Q_l\rangle} $  do not depend on momenta,  
we can integrate over $\overline{P}$, $\widetilde{P}$, $\overline{\xi}$, $\widetilde{\xi}$, so  variables 
$\overline{\xi}$ and $\tilde{\xi}$ have to be equal to zero due to arising delta functions. 
For an isotropic system after averaging over orientations of $k$ we have 
\begin{multline}
	%\begin{eqnarray}	
	S(|k|, \omega) 
	= \frac{8 \pi^3	}{(2\pi)^{6N} N Z^2} 
	\int\hspace{-3pt} {\rm d}\overline{Q} \, {\rm d}\widetilde{Q} 
	%	\nonumber\\&&  	
	\int \hspace{-3pt} {\rm d} t  e^{-{\rm i} \omega t } \times \\
	\hspace{-8mm}\left\langle \overline{Q}   \left|e^{{\rm i}\hat{H}t_{c}}\right|
	\widetilde{Q}  \right\rangle  
	%	   \nonumber\\&&  
	\left\langle \widetilde{Q} \left|e^{-{\rm i} \hat{H}t^*_{c}}\right|
	\overline{Q} \right\rangle  
	%	\nonumber\\&&  	
	\sum_{l} \frac{\sin (|k||\overline{Q}_l-\widetilde{Q}_l|)}{|k||\overline{Q}_l-\widetilde{Q}_l|}.
	\label{srho}	
	%\end{eqnarray}
\end{multline} 
Let us show that from physical point of view the introduced here  the ESCF describes  
scattering  by entangled pairs of electrons at $\overline{Q}_l$ and   $\widetilde{Q}_l $ 
in direction $\iota$ of the monochromatic plane waves with unit amplitude propagating in direction 
$\iota_0$ ($\iota$ and $\iota_0$ are unit vectors). 
Indeed according to Eq.~(\ref{CFA2}) the ESCF is proportional to the sum of products 
$\sum_{1 \leq l \leq N } e^{ {\rm i}\overline{\phi}_l }  e^{{-\rm i}\widetilde{\phi}_l}$.  
From elementary geometric considerations it follows that 
$\overline{\phi}_l - \widetilde{\phi}_l = \frac{2\pi}{\lambda}((\iota-\iota_0)\cdot (\overline{Q}_l- \widetilde{Q}_l) =
\frac{4\pi}{\lambda} sin(\vartheta/2)| \overline{Q}_l- \widetilde{Q}_l |cos(\theta_{l})$,  
where $cos(\vartheta)=(\iota \cdot \iota_0)$ and $\theta_{l}$ is angle between vectors $\overline{Q}_l- \widetilde{Q}_l $ and $\iota-\iota_0$. 
So  averaged over  orientation of vector $\iota$ the $S \sim \mbox{Im} \{\, \overline{e^{ {\rm i}\tilde{k}|\overline{Q}_l- \widetilde{Q}_l |cos(\theta_{l})}}\}$ 
as function versus  $|\overline{Q}_l- \widetilde{Q}_l |$ 
can have peaks  due to the $2 \pi$ periodicity of image exponent. 
Here according to Eq.~(\ref{CFA2})  the ESCF describes scattering and interference between different pairs \cite{sturm1993dynamic}. 
The ESCF is function describing the dynamic scattering properties of unperturbed by the monochromatic plane waves system \cite{sturm1993dynamic}, 
which may be important for experimental diagnostics.  

\section{Heisenberg--Wigner path integral representation and  ``replica trick'' } 
The primary problem of calculation ESCF is that the operators of kinetic $\hat K$   and potential $\hat U$   energies  do not commute; hence, an exact explicit 
analytical formulation for the matrix elements remains unknown.  Nonetheless, this problem can be solved utilizing 
a path integral methodology \cite{feynmanquantum} predicated on the semi-group condition 
$\exp(\kappa\hat{H}) = \exp(\kappa\hat{H}/M) \times \dots \times \exp(\kappa\hat{H}/M)$ $(\kappa = t_{\rm c}^{*}/M$, 
for any positive integer $M$ ). 
For a better understanding of  Eq.~\ref{CFA2} the mathematical structures of  path integral representation of propagators  
$\left\langle \overline{Q}+\frac{\overline{\xi}}{2} 
\left|e^{{\rm i}\hat{H}t_{\rm c}}\right|
\widetilde{Q}-\frac{\widetilde{\xi}}{2}  \right\rangle 	
\left\langle \widetilde{Q}+\frac{\widetilde{\xi}}{2} \left|e^{-{\rm i} \hat{H}t^*_{\rm c}}\right|
\overline{Q}-\frac{\overline{\xi}}{2}  \right\rangle$
is  illustrated in symbolic form by Fig.~\ref{CFA1}  at $\overline\xi=0$ and $\tilde{\xi}=0 $ for simplicity.

\begin{figure}[htp] 
	\centering
	%    \hspace{-1cm}
	\includegraphics[width=0.5\columnwidth,clip=true]{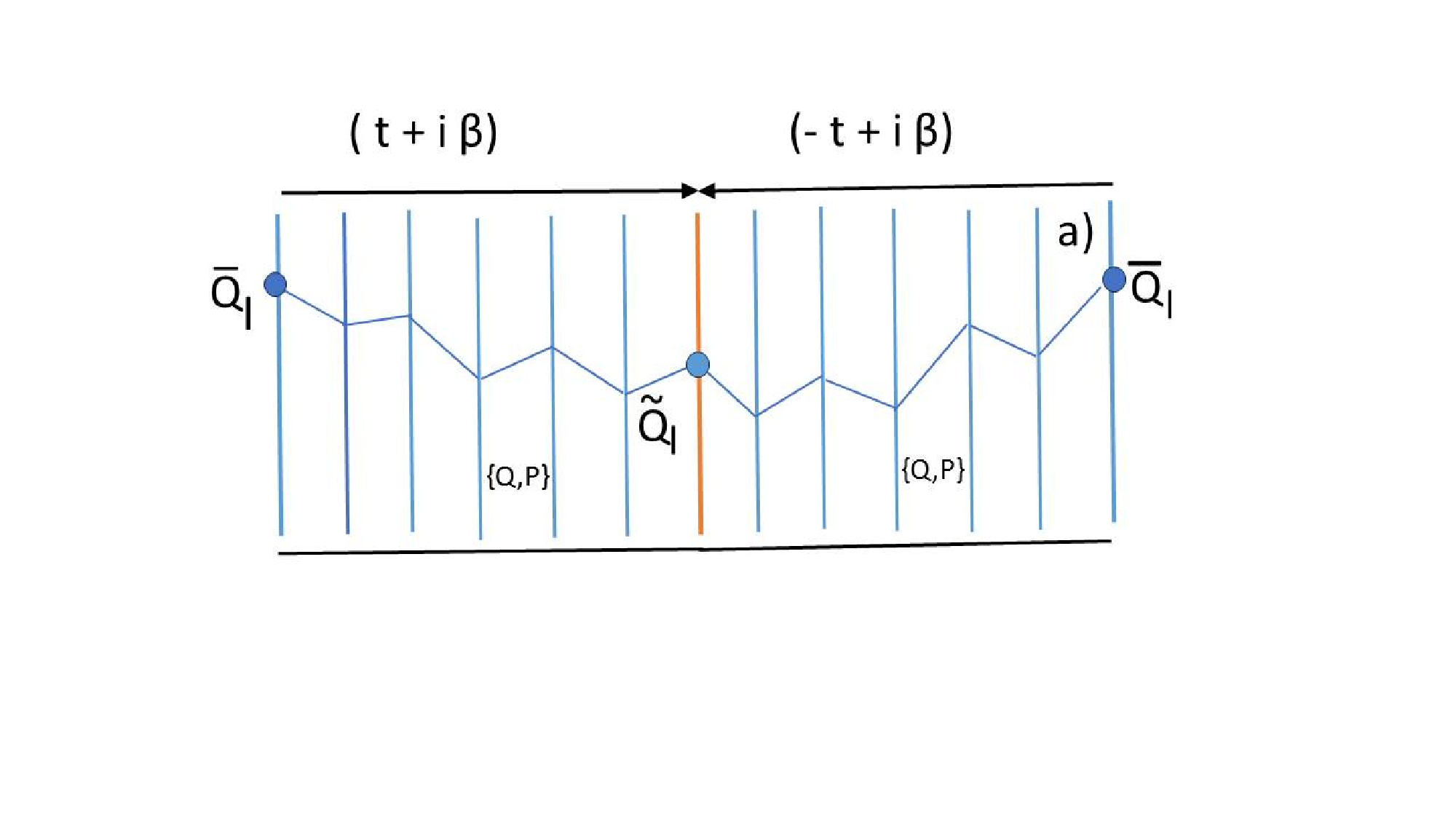}
	\vspace{-0.5cm}    
	\includegraphics[width=0.4\columnwidth,clip=true]{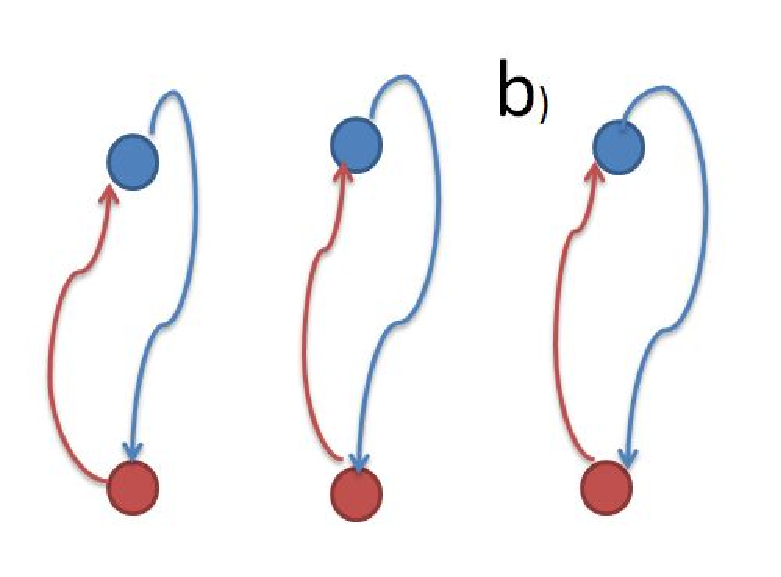}			
	\vskip 0.3cm
	\caption{(Color online) Panel a). 
		A schematic illustration of paths  describing entangled pairs of electrons in path integral representation of the propagators 
		$e^{ {\rm i}\hat{H}t_{\rm c}}$ and $ e^{{-\rm i}\hat{H}t_{\rm c}^{*}}  $ in Heisenberg --Wigner representation. 
		The phase space $\{Q,P\}$ of the many-- body system is represented by the vertical lines with periodic left and right 
		boundary conditions at initial (bar) and final (tilde) ``times''. 
		Generalizing ideas of  \cite{hornberger2009introduction,melko2010finite,humeniuk2012quantum,calabrese2004entanglement,buividovich2008numerical} 
		we describe the ``time of evolution'' 
		by the Heisenberg representation of the Fourier transform of the density operator 	using $t_c= t + {\rm i}\beta$. 
		The ``time'' is  illustrated  by the  horizontal axes. 
		Panel b). Symbolic example of the three entangled pairs in the phase space.  
		\label{correlationc}
	}
\end{figure}  

\section{Generalization of the Wiener--Khinchin theorem} 
Paths in path integral representation of $S(|k|, \omega)$ 
can be considered as a stationary random processes with a generalized distribution defined by propagators \\
$\left\langle \overline{Q}  \left|e^{{\rm i}\hat{H}t_{\rm c}}\right|\widetilde{Q} \right\rangle  	
\left\langle \widetilde{Q} \left|e^{-{\rm i} \hat{H}t^*_{\rm c}}\right| \overline{Q}  \right\rangle$. 
According to the cross-cor\-re\-lation theorem or as well the Wiener--Khinchin theorem  \cite{wiener1930generalized,khinchin1938theory,ziemer2006principles,strichartz2003guide}, 
for a stationary random process 
the expected value of the Fourier transform of a corresponding correlation function can be expressed through  
the expected value of its spectral density. We will use the generalization of this theorem to consider 
the path integral representation of $S(|k|,\omega)$.   

We consider the Fourier transform of the equal time propagators in the expression for $S(|k|,\omega)$  
(see Eq.~(\ref{CFA1}),  Fig.~1  and Eq.~(\ref{srho})). 
By representing each propagator at the same time as the inverse Fourier transform in Eq.~(\ref{srho}), we obtain:

\begin{multline}
	\tilde{F}\left(\overline{Q\xi };\widetilde{Q\xi } ;\omega;  \right)  = 
	\int  {\rm d}  \tilde{t} \,	 e^{-{\rm i} \omega  \tilde{t}  } 
	\left\langle \overline{Q}+\frac{\overline{\xi}}{2}   \left|e^{{\rm i}\hat{H}\tilde{t}_{\rm c}}\right|
	\widetilde{Q}-\frac{\widetilde{\xi}}{2} \right\rangle  
	\left\langle \widetilde{Q}+\frac{\widetilde{\xi}}{2}  
	\left|e^{-{\rm i} \hat{H}\tilde{t}^*_{\rm c}}\right|
	\overline{Q}-\frac{\overline{\xi}}{2}  \right\rangle  \\
	=\int  {\rm d} \tilde{t} \,	 e^{-{\rm i} \omega  \tilde{t} } 
	\frac{1}{2 \pi}\int {\rm d} \omega_1 \, F^+\left(\overline{Q\xi };\widetilde{Q\xi } ;\omega_1 \right) e^{{\rm i} \omega_1 \tilde{ t}}  %\\ \times 
	\frac{1}{2 \pi}\int {\rm d} \omega_2 \, F^-\left(\widetilde{Q\xi };\overline{Q\xi } ;\omega_2 \right) e^{{\rm i} \omega_2 \tilde{ t}}  \\  
	{} = \frac{1}{4 \pi}\int {\rm d} \omega_+ \,\int {\rm d} \omega_- \, \delta(\omega-\omega_+)) %\\ \times 
	F^+\left(\overline{Q\xi };\widetilde{Q\xi }
	;(\omega_+ + \omega_-)/2 \right) 
	F^-\left(\widetilde{Q\xi };\overline{Q\xi } ; (\omega_+ - \omega_-)/2 \right), \\
	= \frac{1}{4 \pi}   \int {\rm d} \omega_- \,F^+\left(\overline{Q\xi };\widetilde{Q\xi } ;(\omega + \omega_-)/2 \right) %\\ \times 
	F^-\left(\widetilde{Q\xi };\overline{Q\xi } ; (\omega - \omega_-)/2 \right), 
\end{multline}
%\end{multline}
where 
\begin{multline}
	%\begin{eqnarray}
	F^+\left(\overline{Q\xi};\widetilde{Q\xi} ;\omega \right)=
	\int \hspace{-3pt} {\rm d}   t \left\langle \overline{Q}  \left|e^{{\rm i}\hat{H}t_{\rm c}}\right|\widetilde{Q} \right\rangle   
	e^{-{\rm i} \omega   t},  %\\
	%\nonumber
	F^-\left(\widetilde{Q\xi};\overline{Q\xi} ;\omega \right) = 
	\int \hspace{-3pt} {\rm d}   t \left\langle \widetilde{Q} \left|e^{-{\rm i} \hat{H}t^*_{\rm c}}\right| \overline{Q}  \right\rangle 
	e^{-{\rm i} \omega   t} 
	\nonumber
	%	\label{dt6}
\end{multline} 
%\end{eqnarray}	
%where  
and $\omega_1=(\omega_+ + \omega_-)/2$, $\omega_2=(\omega_+ - \omega_-)/2$, the Jacobian 
of the transformation $(\omega_1, \omega_2)\to (\omega_+, \omega_-)$ is 1/2 and we used 
the integral representation of 
the $\delta$-function 
$\int \text{d}\Delta\tilde{t} e^{-i\Delta\tilde{t}(\omega - \omega_1 -\omega_2)} = 2\pi\delta(\omega - \omega_+)$.  

\textbf{To clarify mathematical and physical sense of this generalization  } 
let us consider Wigner representation of $F^+$ and $F^-$ through the path integrals and Hamiltonian 
functions in phase space once more, so we obtain (see Supplemental materials): 
\begin{multline}	
	\tilde{F}(\overline{Q\xi }, \widetilde{Q\xi }, \omega) = 
	\frac{1}{4 \pi}\int {\rm d} \omega_+ \, \int {\rm d} \omega_- \, \delta(\omega-\omega_+) %\\ \times
	F^+\left(\overline{Q\xi };\widetilde{Q\xi } ;(\omega_+ + \omega_-)/2 \right) \\ \times
	F^-\left(\widetilde{Q\xi };\overline{Q\xi } ; (\omega_+ - \omega_-)/2 \right) 
	\\ 
	=\frac{1}{4 \pi} 
	%	\nonumber\\&& 	
	2\pi \left(\frac{1}{2\pi}\right)^{3NM} \int {\rm d} \omega_- \, %\\ \times
	\int \hspace{-3pt} {\rm d} P
	\delta ((\omega + \omega_-)/2 -H(P,Q)) \,	W^+\left(P,Q\right) \\
	%\delta(\omega-\omega_+))  
	{} \times  2\pi  \left(\frac{1}{2\pi}\right)^{3NM}  %\\ \times
	\int \hspace{-3pt} {\rm d} P^{\prime}
	\delta ((\omega - \omega_-)/2 -H(P^{\prime},Q^{\prime})) \,	W^-\left(P^{\prime},Q^{\prime}\right).
	\\
	=2\pi  \left(\frac{1}{2\pi}\right)^{6NM}  %\\ \times
	\int \hspace{-3pt} {\rm d} P {\rm d} P^{\prime}
	\delta (\omega-(H(P,Q)+H(P^{\prime},Q^{\prime})))  \, %\\ \times
	W^+\left(P,Q\right) W^-\left(P^{\prime},Q^{\prime}\right),
	\label{Ftilde}	
	%	\label{GGGFFF}
\end{multline}
%\end{multline}
where  the sum $H(P,Q)+H(P^{\prime},Q^{\prime})$ is the Hamilton function of the whole  phase space path (see Fig.~1) and 
we use the convolution of delta--functions $\int\delta(x - y)\delta(-x -z)dx = \delta(y + z)$. 

Here we expressed the $F^+\left(\overline{Q\xi};\widetilde{Q\xi} ;\omega \right)$ and $F^-\left(\widetilde{Q\xi};\overline{Q\xi} ;\omega \right)$  
through the inverse Fourier transform of the Weyl symbols of $\hat{H}$ being the Hamilton function $H(P,Q)$ and get  
the expressions  $F^{\pm} \sim \int \text{d}P\delta(\omega - H(P, Q)) W^{\pm}(P, Q)$. 
Here $H(P,Q)=\sum_{j=1}^{M} H(P_j,Q_j)/M$ is the Hamiltonian function 
of the phase space trajectory, described by the vector set $\{ Q,P \} $; 
$Q=\{Q_1 , \dots , Q_{M}\}$,  $P=\{P_1 , \dots , P_{M}\}$ and 
$H(P_j,Q_j)= \left\langle P_j|P_j \right\rangle/2m + U(Q_j) $ is  the Hamiltonian function  at a given $j$ 
(see Figure~\ref{correlationc} ).

The $S(|k|, \omega)$  
is defined by the spectral densities $F^+F^-$ and we can use the probability 
corresponding to $W=W^+W^-$ in WPIMC simulations to generate random paths in the phase space. 
So finally the $S(|k|, \omega)$ is :
\begin{multline}
	%\begin{multline} 
	%\begin{eqnarray}
	S(|k|, \omega)  
	= \frac{16 \pi^3	}{(2\pi)^{6N} N Z^2 }  
	\int\hspace{-3pt} {\rm d}\overline{Q} \, {\rm d}\widetilde{Q}   \, 
	\tilde{F}\left(\overline{Q};\widetilde{Q} ;\omega;  \right) \times {} %\\
	\sum_l \frac{\sin (|k||\overline{Q}_l-\widetilde{Q}_l|)}{|k||\overline{Q}_l-\widetilde{Q}_l|} \\
	= \frac{16 \pi^3	}{(2\pi)^{6N} N Z^2 }  
	\int\hspace{-3pt} {\rm d}\overline{Q} \, {\rm d}\widetilde{Q} \times {} %\\
	\frac{1}{4 \pi}\int {\rm d} \omega_+ \, \int {\rm d} \omega_- \, \delta(\omega-\omega_+) \times {}\\
	F^+\left(\overline{Q\xi };\widetilde{Q\xi } ;(\omega_+ + \omega_-)/2 \right) \times {} %\\ 
	F^-\left(\widetilde{Q\xi };\overline{Q\xi } ; (\omega_+ - \omega_-)/2 \right) 
	\sum_l \frac{\sin (|k||\overline{Q}_l-\widetilde{Q}_l|)}{|k||\overline{Q}_l-\widetilde{Q}_l|}. 	
	\label{sskk}
	%\end{multline}	
\end{multline}

Thus, $S(|k|,\omega)$) can be obtained via a WPIMC simulation
of random trajectories in the phase space  with the probability defined by the product $W^{\pm}(P, Q)$ 
as $F \sim \int \text{d}P\delta(\omega - H(P, Q)) W(P, Q)$ (see Eq.~(\ref{Ftilde}). 
These virtual trajectories (see Fig.~1  and Eq.~(\ref{srho})).  
are non- classical and do not satisfy  Hamiltonian equations . 
Calculations of the Hamiltonian function in delta functional $\delta(\omega -H(P,Q)) $ 
for each virtual trajectoriy allows to calculate averaged over trajectories  \textbf{histogram} of the values  
$\frac{1}{N}\sum_l \frac{\sin (|k||\overline{Q}_l-\widetilde{Q}_l|)}{|k||\overline{Q}_l-\widetilde{Q}_l|}$  
and the weight functional $\delta (\omega-H(P,Q))$ as function of  $\omega $).

To calculate $ S(|k|,\omega)$) 
the Markovian chains of particle configurations 
were generated in a Monte Carlo cell with periodical topology 
according to \cite{hornberger2009introduction,melko2010finite,humeniuk2012quantum}
using the Metropolis algorithm 
(see Supplemental material for  details)). 
From $10^6$ to $3\times 10^6$ equilibrium configurations of 300, 600 and 900 particles represented 
by twenty and forty ``beads'' were used to calculate average values. 
The convergence and statistical error of the calculated functions were tested by increasing the number of particles and beads 
as well as the length of the Markovian chain. 
The statistical error of the WPIMC calculations practically coincides with the amplitude of small oscillations 
of the functions presented in Figures below.  

%%%%%%%%%%%%%%%%%%%%%%%%%%%%%%%%%%%%%%%%%%%%%%%%%%%%%%%%%%%%%%%%%%%%%%%%%

\section{\label{sec:results}Results of WPIMC simulations}\label{simulations}   

Within the ``replica trick'' approach \cite{hornberger2009introduction,melko2010finite,humeniuk2012quantum}) 
let us consider quantum entangled electron pairs within 
%  and uncorrelated positive charges within 
a modified UEG model being a two-component Coulomb system. 
The electroneutrality condition should be satisfied  
%with all interparticle interactions 
as otherwise the thermodynamic limit does not exist and the so-called ``Coulomb explosion'' takes place. 
In the considered UEG model the positive charge component was treated as 
a gas of particles \textit{without any correlations}  simulating the neutralizing background. 
Hamiltonian function $H=K+U^c$ of this UEG model 
% with the Hamiltonian  ${\hat H}={\hat K}+{\hat U}^c$ 
contains the kinetic energy $K$ and interaction energy of positive charges and entangled electron pairs  
$U^{c} = U_{pp}^c + U^c_{ee} + U^c_{ep}$, where  ${\hat U}_{pp}^c$ and ${\hat U}_{ep}^c$ are a contributioss 
of the positive background  itself (uncorrelated positive charges) and its interaction with electrons have to be 
accounted for in calculation $H(P,Q)$ in $\delta (\omega-H(P,Q))$ in the WPIMC averaged histogram (see above).
%For uncorrelated positive charges  in considered model 
%we have to put ${\hat U}_{pp}^c \equiv 0$ and ${\hat U}_{ep}^c \equiv 0$ in the Hamiltonian  of the density matrix and time propagators, 
Below density of UEG is characterized by 
the Brueckner parameter $r_s=a/a_B$ 
defined as the ratio of the mean distance between electrons $a=\left[3/(4 \pi \tilde{\rho}) \right]^{1/3}$ 
to the Bohr radius $a_B$. 

\subsection{The spin--resolved radial distribution functions}\label{rrdf}

Let us start considering the WPIMC results from the detailed discussion  of  physical  peculiarities of the modified UEG model.  
In the canonical ensemble under Boltzmann statistics in PIMC simulations of the system of particles  there are  four 
different quantum radial distribution functions (RDF) \cite{sese2002structural} describing the spatial correlations between:  
%, depending on what is correlated: 
beads center of mass, beads at same time or beads at different times 
as well as intranecklace bead correlations. 
The first three radial distribution functions tend to the same classical limit without quantum spreading.

For a degenerated system of particles in the Wigner approach to quantum mechanics in the phase space 
(see Eq.~(\ref{sskk}) ) it is natural to consider the RDF $g_{ab}$ as   
\begin{eqnarray}  \label{gab-rho}
	&&g_{ab}(r) =  \int {\rm d} \bar{Q}  \, \delta(|\bar{Q}^a_1-\bar{Q}^b_1|-r)\, 
	\nonumber\\&& \times 
	\int {\rm d} \tilde{Q} \, {\rm d} \bar{P}\, {\rm d} \tilde{P} \, 
	W^+(\overline{PQ})\,W^-(\widetilde{PQ}) 
\end{eqnarray}
where $\bar{Q}^a_1$ and $\bar{Q}^b_1$ are coordinates, while $a$ and $b$ label unentangled electrons with the spin variables 
or uncorrelated positive charges with number one (see Eq.~(\ref{CFA2})). 

Let us analyze the RDFs of electrons and positive uncorrelated neutralizing charges. 
As it follows from the Wigner approach to quantum dynamics 
the RDF $g_{ee}$ is proportional to the probability density to find a pair of electrons  of types $a$ and $b$
at a certain distance $r$ from each other. The RDF depends only on the difference of coordinates due to the translational invariance
in an isotropic system. 
In a classical non-interacting system of electrons  $g_{ab} \equiv 1 $, while  
interaction, quantum effects and statistics result in the spacial redistribution of electrons and a non-monotonic RDF. 

\begin{figure}[htp] 
	\includegraphics[width=0.5\columnwidth,clip=true]{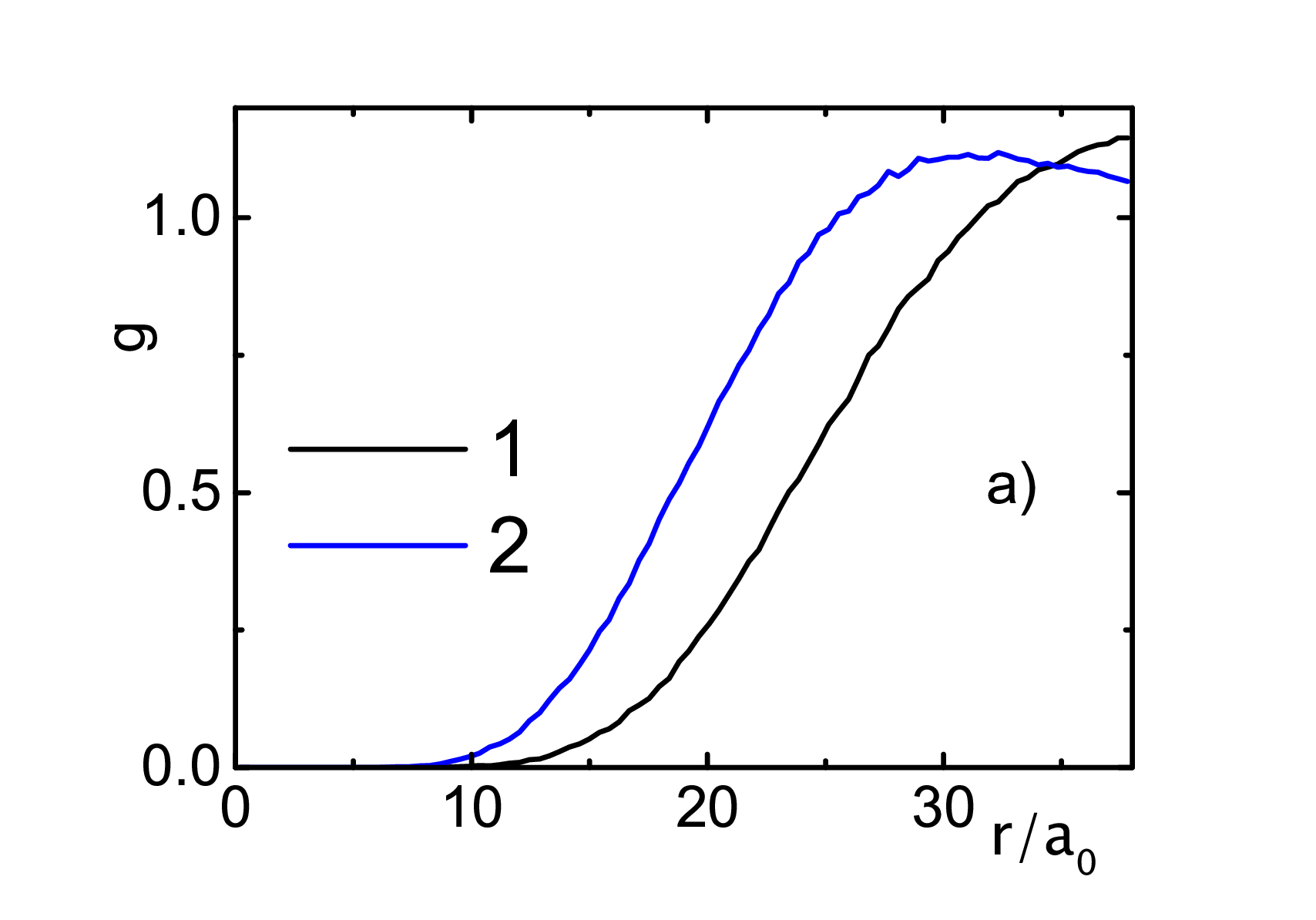}		
	\includegraphics[width=0.5\columnwidth,clip=true]{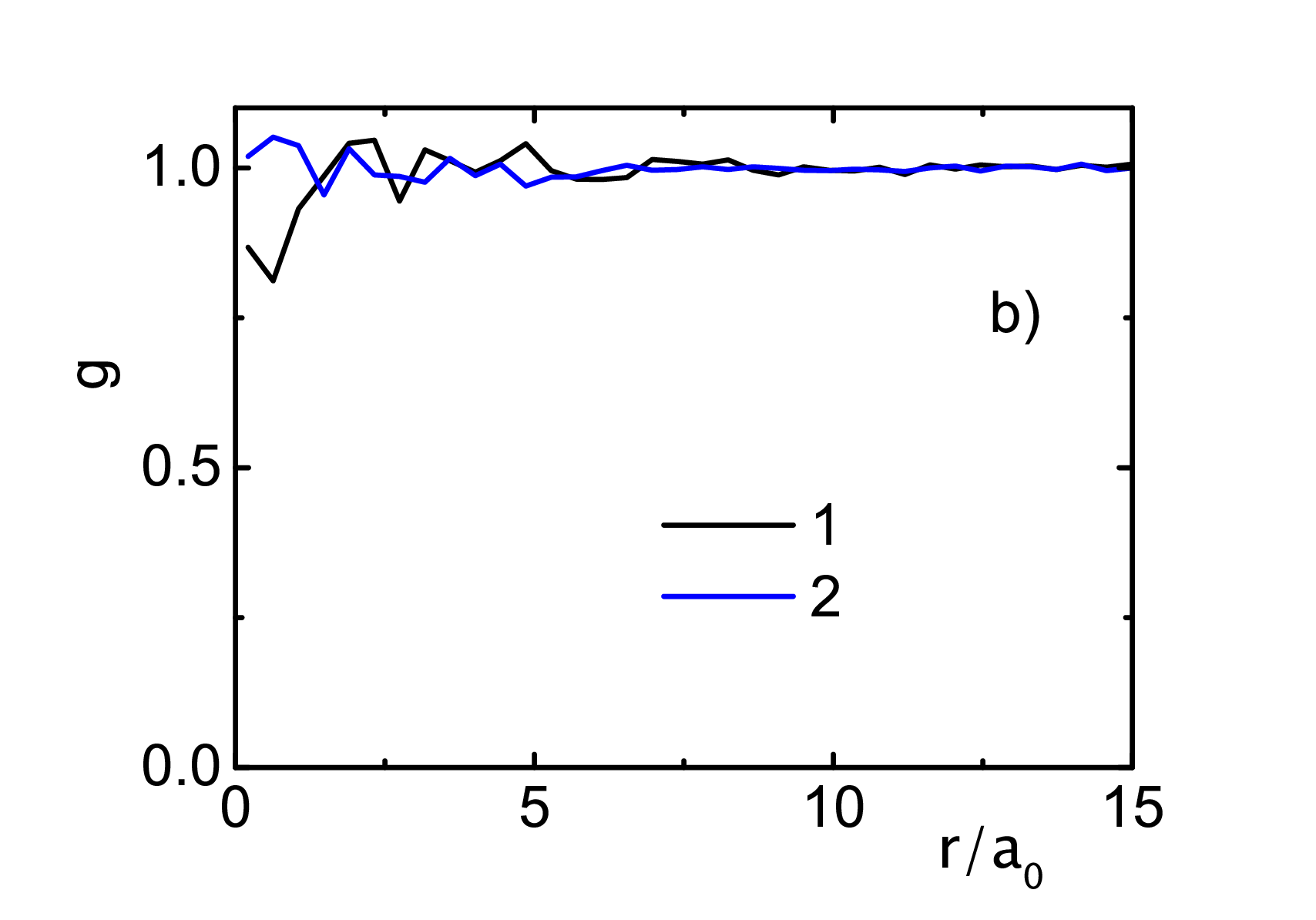}	
	%	%	\vskip 0.3cm
	%	%\hspace{.2cm}
	\caption{(Color online) 
		The WPIMC RDFs at $T=0.005 Ry$ $r_s=20$. 
		Panel a) -- electron--electron RDFs $g_{ee}$.  
		Lines: 1 -- the same spin projections, 2 --the opposite spin projections.  
		Panel b). Lines: 1 -- the electron -- positive charge $g_{ep}$, 2 -- the positive -- positive charges $g_{pp}$. 
		The statistical error of the WPIMC calculations practically coincides with the amplitude of small oscillations. 
		\label{gr1}	}
\end{figure}

Figure~\ref{gr1} shows the WPIMC RDFs for strongly coupled degenerated electrons and positive uncorrelated neutralizing charges.  
At small interparticle distances electron--electron RDFs $g_{ee}$ (lines 1 and 2 on panel a)) tends to zero due to the Coulomb repulsion. 
Additional fermi repulsion of electrons with the same spin projections (line 1) described by the exchange determinant arising due to 
the antisymmetriztion of the density matrix (see for details Supplemental material) 
is the reason of the faster decay in comparison with opposite spin case (line 2) at distances of the order of the thermal wavelength.  
This leads to the formation of cavities and, as consequence, to arising excluded--volume peaks at a distance of $\sim 30a_0 $. 
At larger interparticle distances the RDFs tend monotonically to unity due to the short range of the effective interparticles interaction. 

Figure~\ref{gr1}b) shows practically equal to unity WPIMC RDFs for electron -- positive charge  $g_{ep}$ and uncorrelated 
positive charge--positive charges $g_{pp}$ (lines 1 and 2). 

\begin{figure}[htp] 
	\centering
	\includegraphics[width=0.45\columnwidth,clip=true]{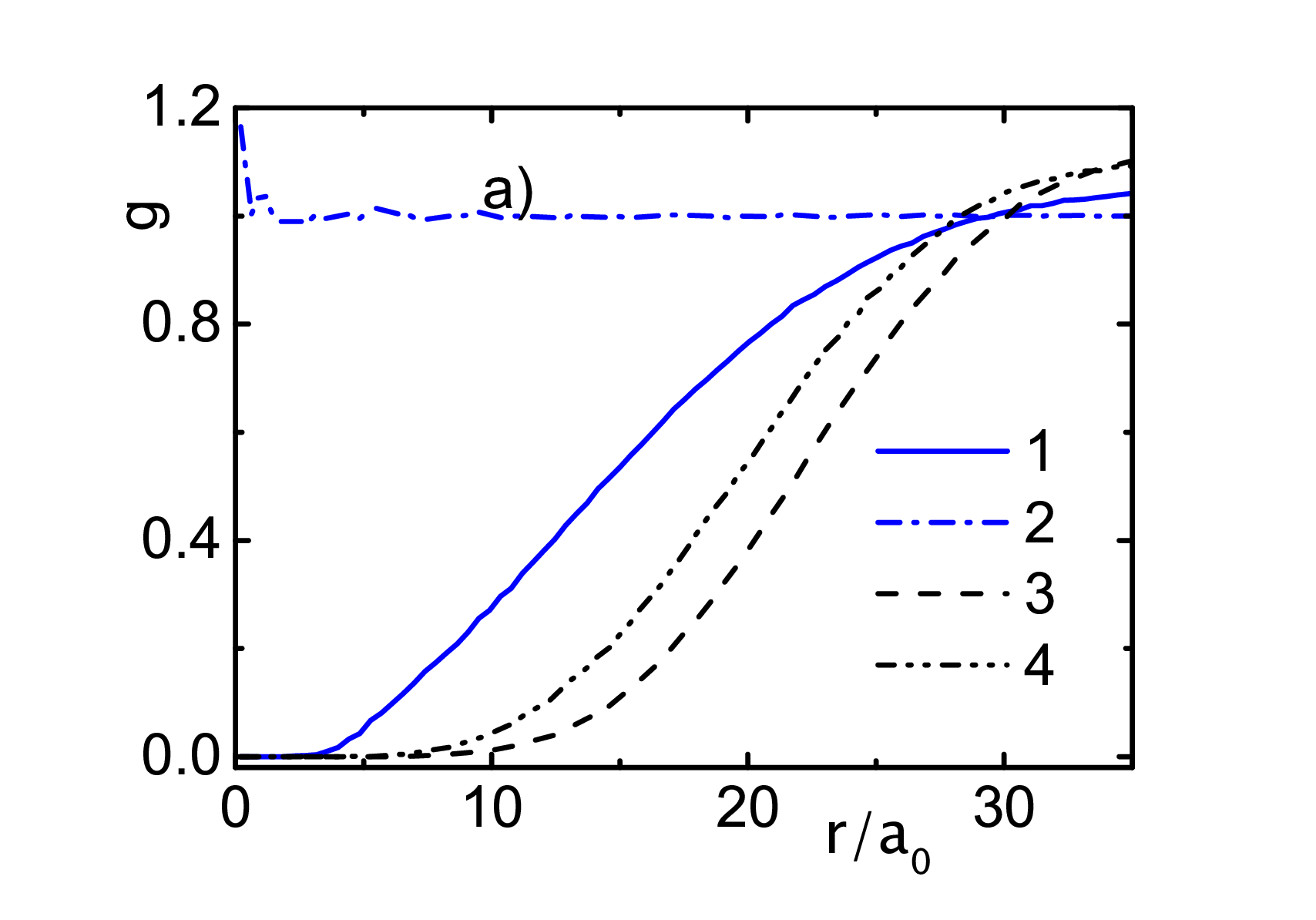}		
	\includegraphics[width=0.45\columnwidth,clip=true]{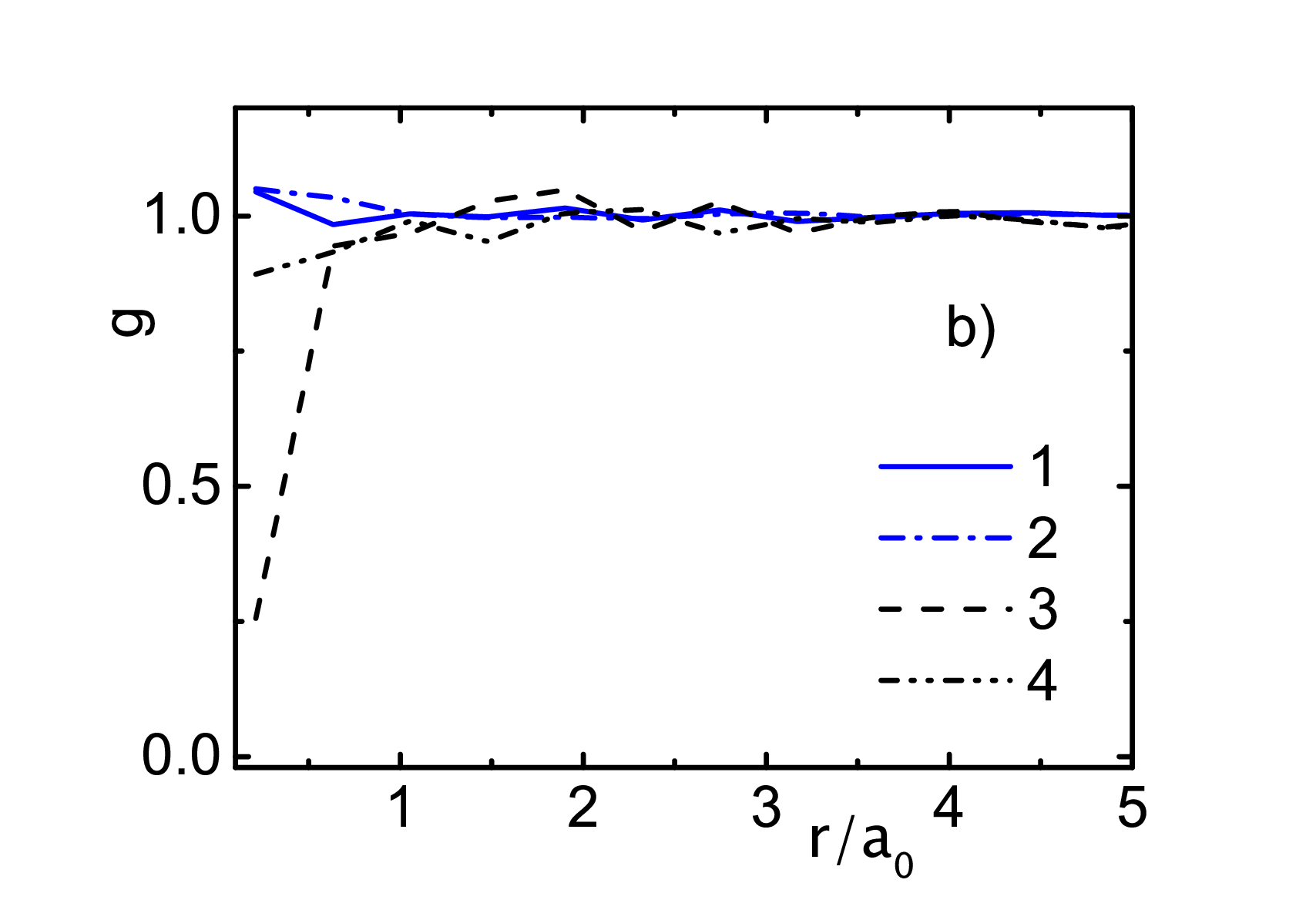}	
	%%	%	%	\vskip 0.3cm
	%%	%	%\hspace{.2cm}
	\caption{(Color online) 
		The WPIMC RDFs at $T=0.01 Ry$ $r_s=20$. 
		Panel a) -- electron--electron $g_{ee}$. 
		Ideal electrons: the same/opposite spin projections -- lines 1/2. 
		Interacting electrons: the same/opposite spin projections -- lines 3/4. 
		Panel b) -- positive--positive charge $g_{pp}$. Lines 1/2 -- ideal/interacting electrons; 
		electron-- positive charge $g_{ep}$ -- lines 3/4 for interacting/ideal  electrons. 
		\label{gr2}	
	}
\end{figure}

Analogously to the Figure~\ref{gr1}a) the Figures~\ref{gr2}a) (lines 3 and 4) present as example  behavior of radial distribution funcions $g_{ee}$ 
at higher temperature but the same density, while lines 1 and 2 describe $g_{ee}(r)$  of noninteracting degenerated electrons 
for the same and opposite spin projections. 

Figures~\ref{gr2}b) present radial distributions  $g_{pp}(r)$ and  $g_{ep}(r)$ 
for ideal (lines $1$ and $2$)  and interacting (lines $3$ and $4$ )) electrons respectively. 
The RDF $g_{pp}(r)$  of the positive uncorrelated  charges uniformly distributed in space   
is identically equal to unity ( Figure~\ref{gr2}b),  lines 1 and 2 ). Small discrepancy in behavior of lines 1 and 2  
additionally illustrates the value of the WPIMC statistically error of RDFs for ideal and interacting electrons. 

Let us note that contrary to RDF $g_{ep}(r)$ on Figure~\ref{gr1}b) (line 1)) the analogous RDF $g_{ep}(r)$ of Figure~\ref{gr2}b) (line 3)  
demonstrates an unexpected significant drop.  The physical reason of this discrepancy is the three-particle effect arising 
due to the electron repulsion preventing for any two electrons 
being in the vicinity of any uncorrelated  positive charge.  

\subsection{Decoherence dynamics of entangled electron pairs, ESCF and time ESCF}\label{rrdf}

\begin{figure}[htp] 
	\centering
	\includegraphics[width=0.32\columnwidth,clip=true]{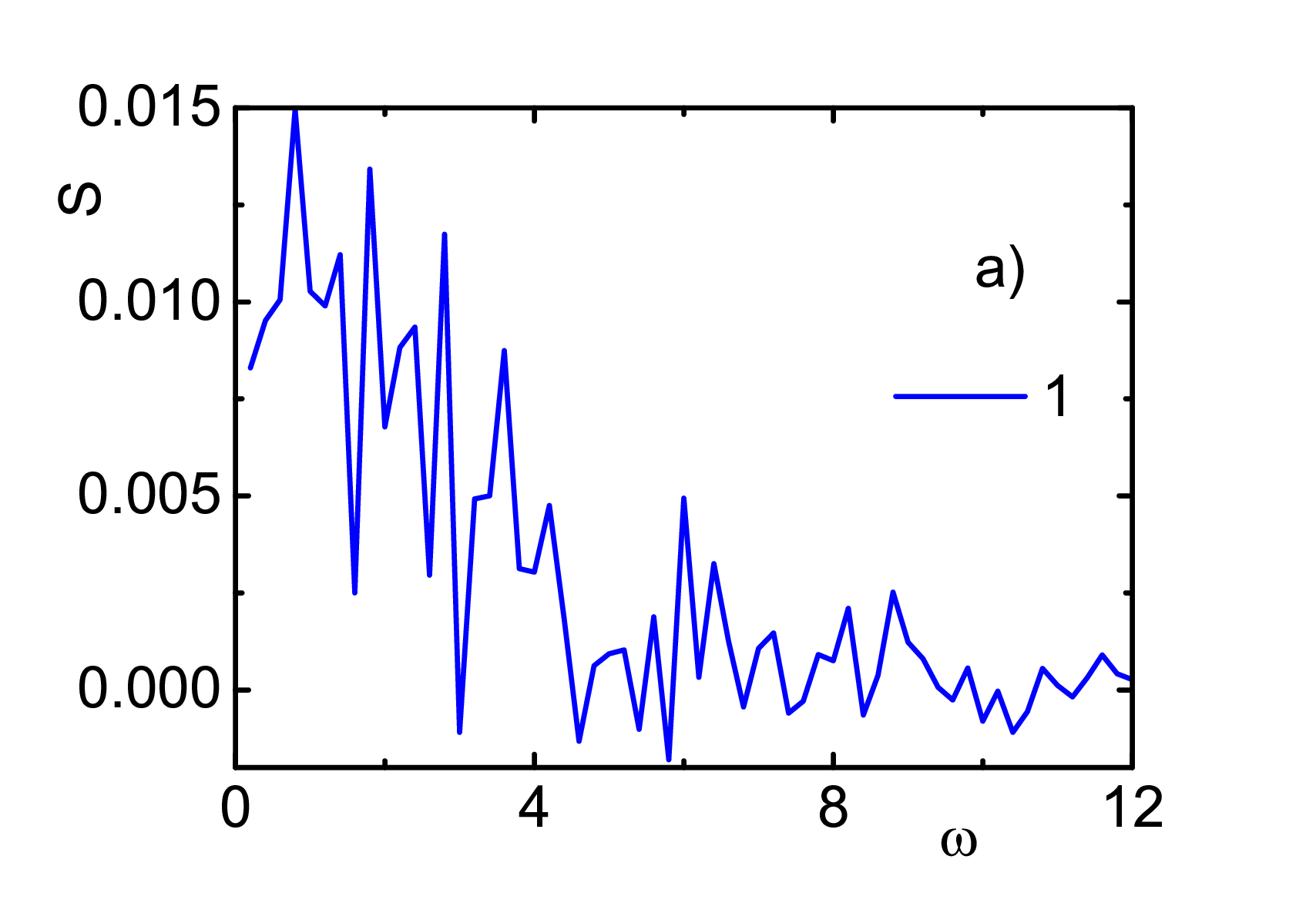}		
	\includegraphics[width=0.32\columnwidth,clip=true]{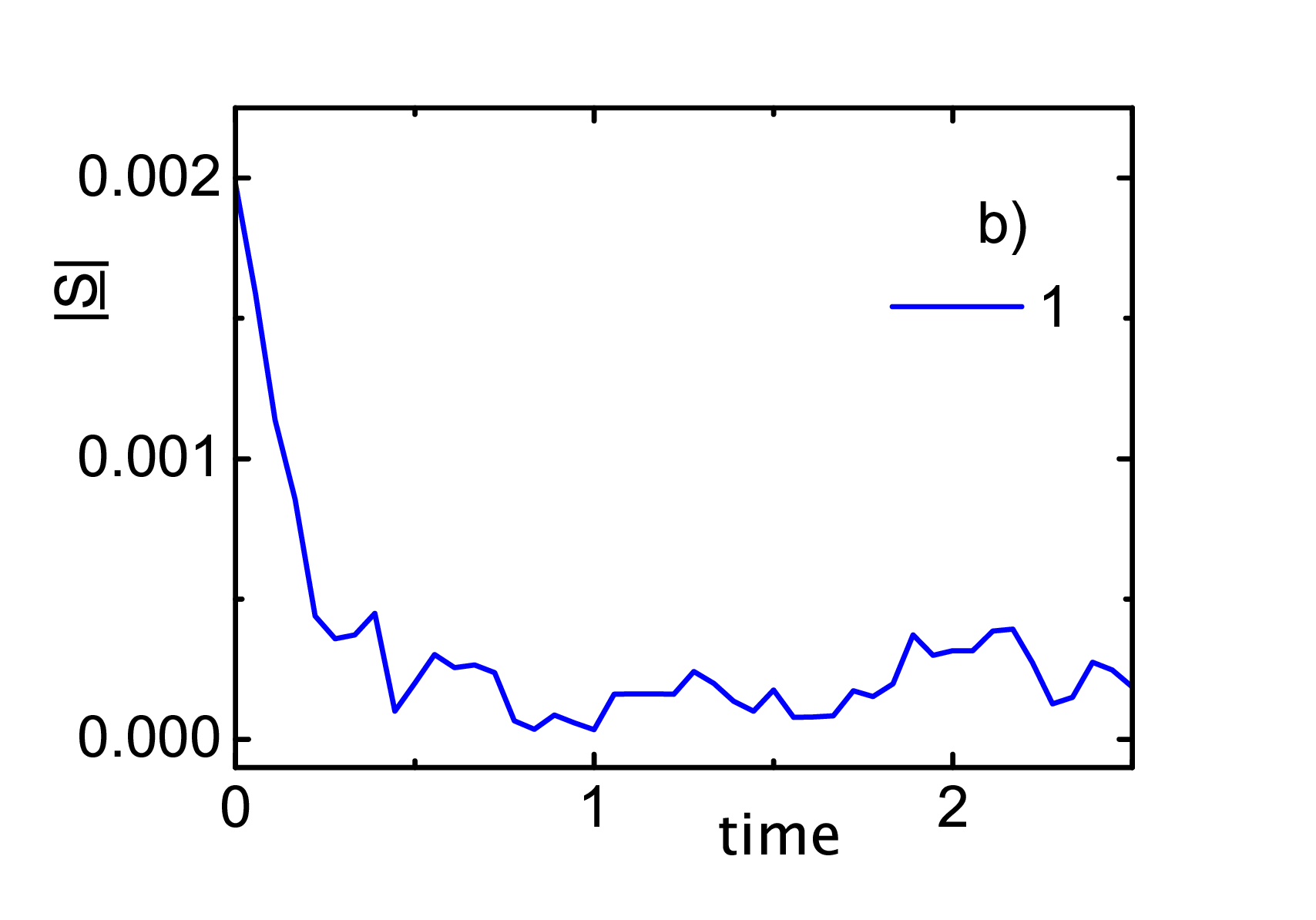}	
	\includegraphics[width=0.32\columnwidth,clip=true]{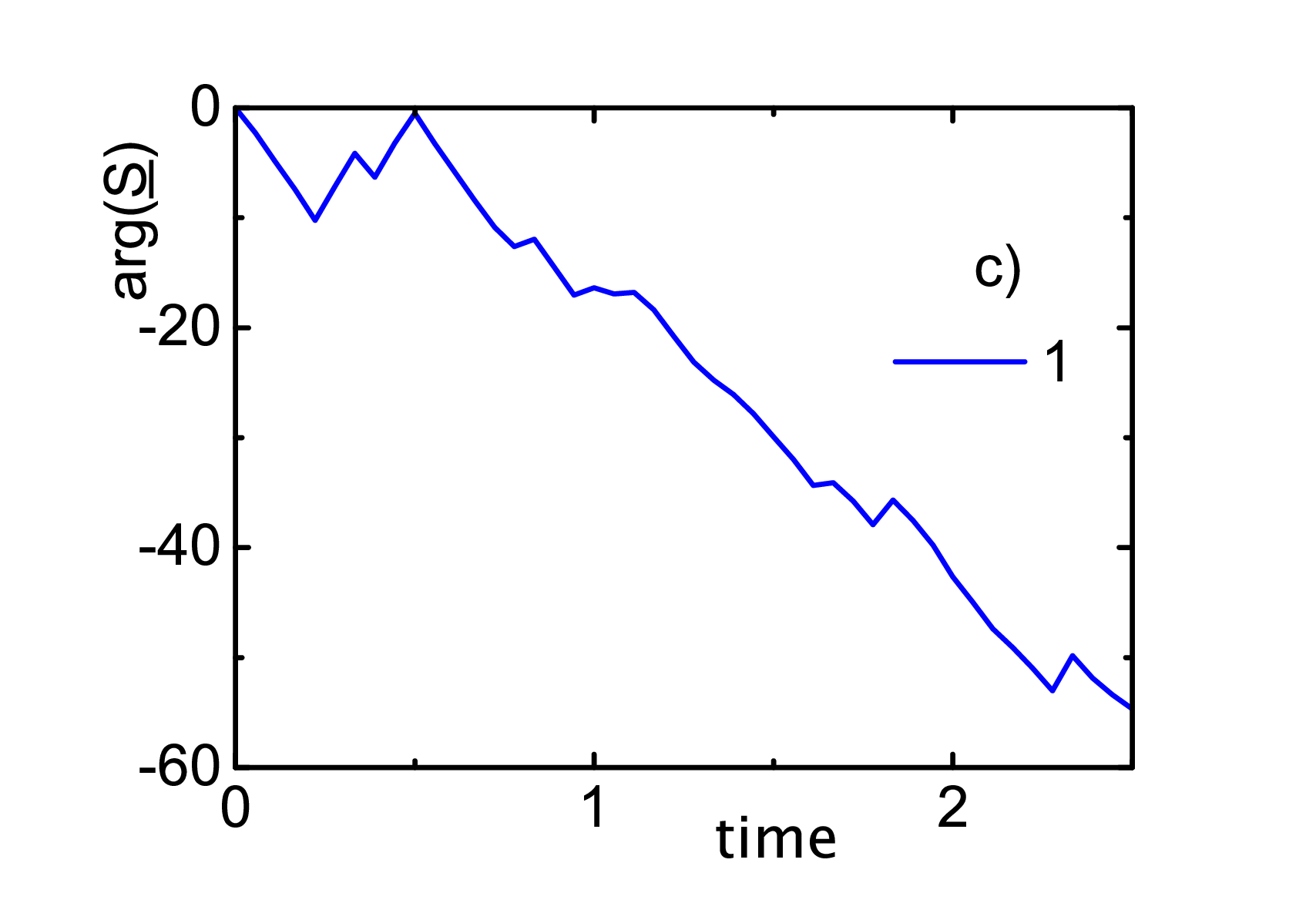}	
	%	%	\vskip 0.3cm
	%	%\hspace{.2cm}
	\caption{(Color online) 
		The $S(|k|, \omega)$ and its inverse Fourier transform $\underline{S}(|k|, t )=|\underline{S}| e^{{\rm i} arg(\underline{S}) }$ at $T =0.005 Ry $ $r_s=20$ and  $|k|=0.1$ 
		in conditional units ( lines $1$). 
		%The statistical error of the WPIMC calculations practically coincides with the amplitude of small oscillations		
		\label{harm}	}
\end{figure}

Figures~\ref{harm} present ESCF and  its inverse Fourier transform -- the time entangled state correlation functions (TESCF) at low temperatures  
for UEG model of entangled electron pairs, which accounts for entangled pairs interactions (factor 1), 
interaction of electrons with positive neutralizing uncorrelated charges (factor 2)
and exchange energy with heat--bath (in canonical ensemble, factor 3).  
Here, as often referred to in literature, a quantum electron in path integral approach is represented by a path of  
``beads, connected via harmonic like springs  to form a Gaussian closed chain'' and called also a ``ring polymer'' 
(see Figures~\ref{correlationc} and  Supplemental material). 
Coupled systems described by the interacting oscillators have been widely used in classical and quantum mechanics 
and appear in various models. 

%\begin{widetext}
%\end{widetext}

Oscillation modes with discrete like spectrum  obtained by WPIMC is presented in Fig.~\ref{harm}a). 
Dynamic properties  of the such  ``ring polymer''  can  qualitatively  resemble  the harmonic spectrum of the  
single-particle or collective oscillations \cite{lievens2006harmonic,lievens2008harmonic}, but here 
with  possible  deformation by degeneracy and interelectron interaction. 

A closed isolated quantum system will always follow unitary quantum dynamics. 
However quantum system interacting with environment is an open quantum one like entangled electron pairs 
coupled to  heat--bath  ( factors 3) ) and uncorrelated positive charges 
simulating a neutralizing background (factor 2). The coupling of a quantum system with environment 
leads to decoherence and decay of the TESRF with certain characteristic decoherence time (see Fig.~\ref{harm}(panel b) and c) ). 
Decoherence is the fundamental mechanism responsible for destroying entanglement and, as consequence, classical transition.  
In fact, decoherence is one of the main obstacles for implementation of multi-qubit 
entangled states (for example see review in \cite{lievens2006harmonic,lievens2008harmonic,valiev2005quantum,zalivako2025quantum}.  

\begin{figure}[!t] 
	\centering
	\includegraphics[width=0.32\columnwidth,clip=true]{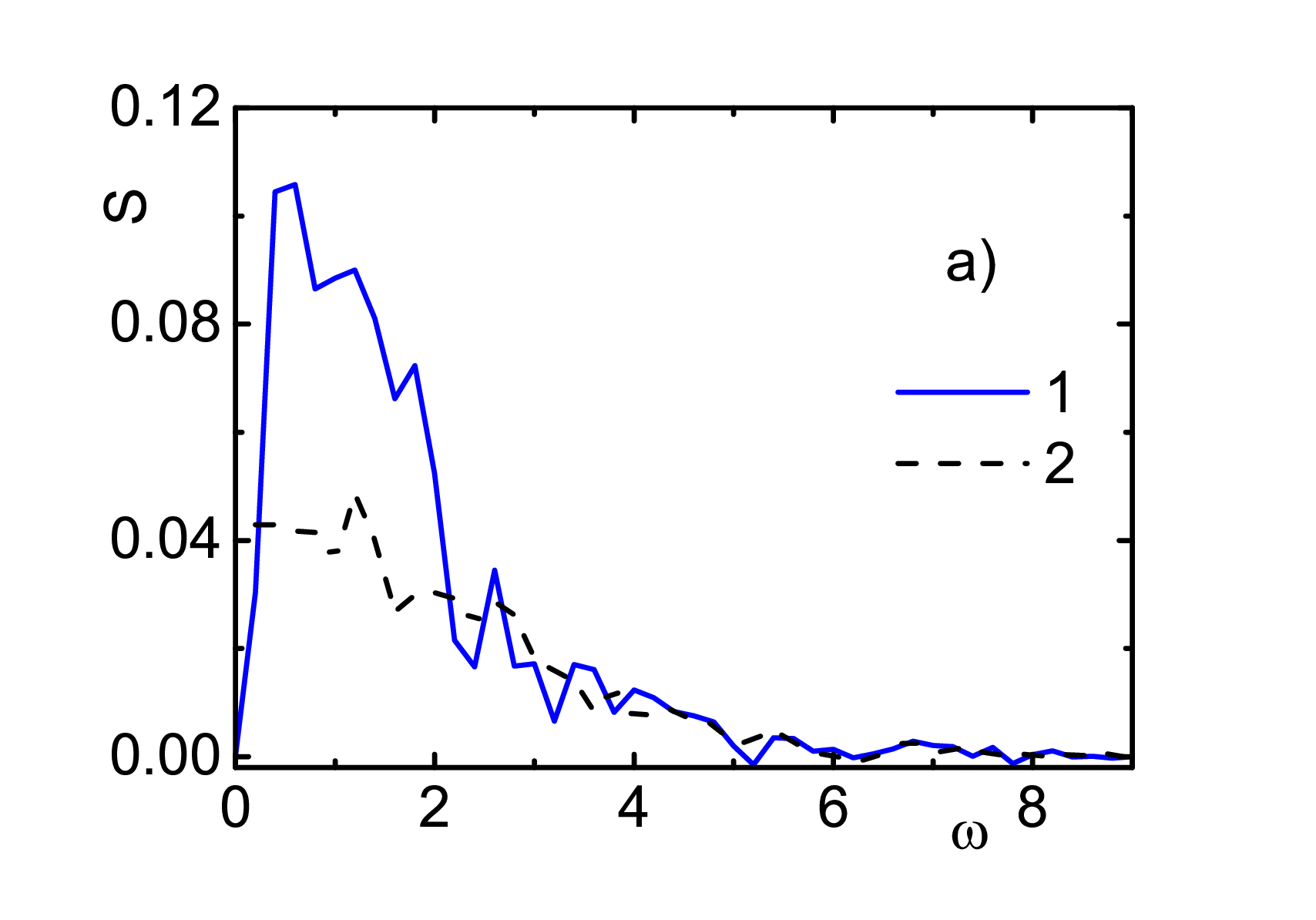}		
	\includegraphics[width=0.32\columnwidth,clip=true]{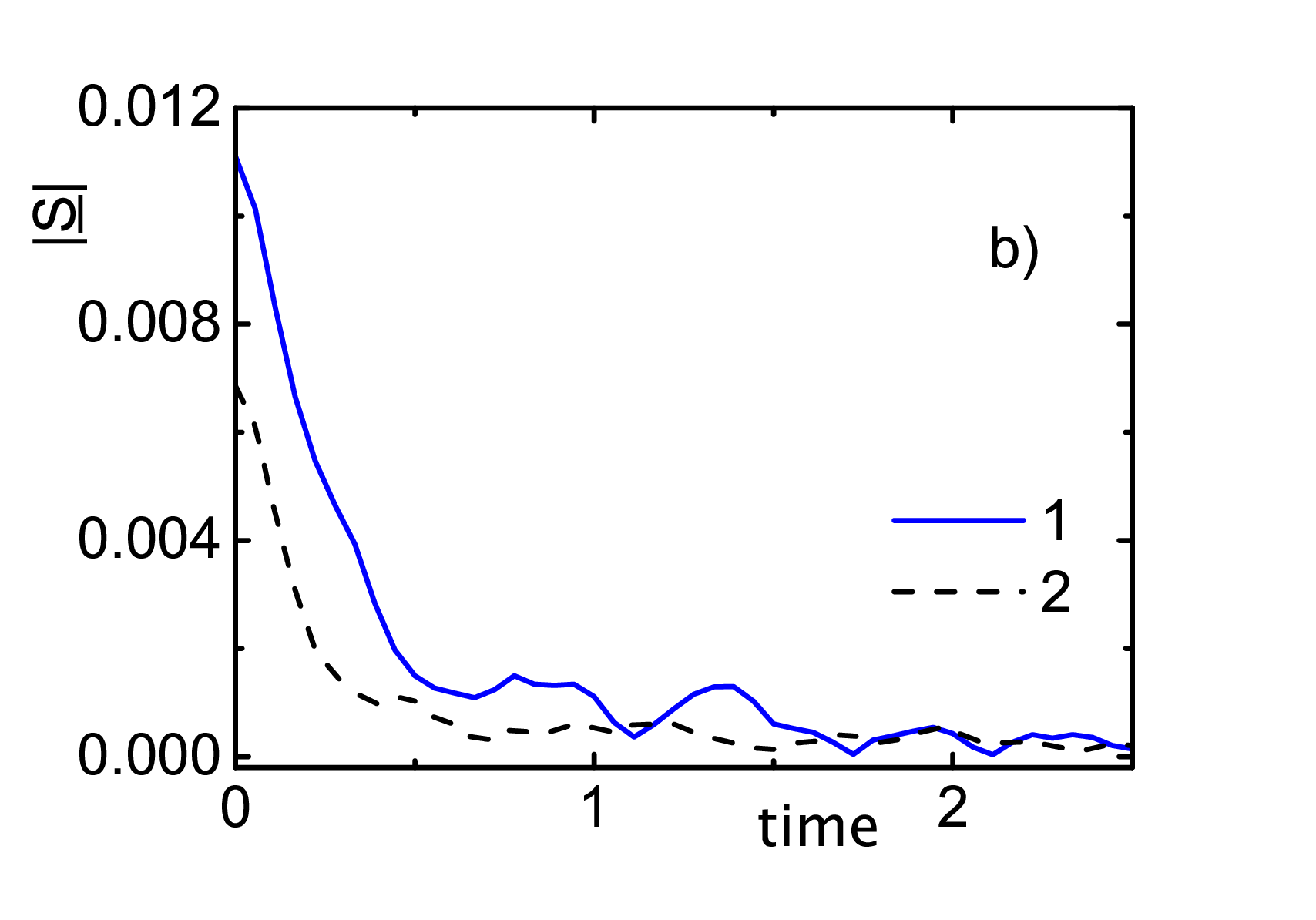}	
	\includegraphics[width=0.32\columnwidth,clip=true]{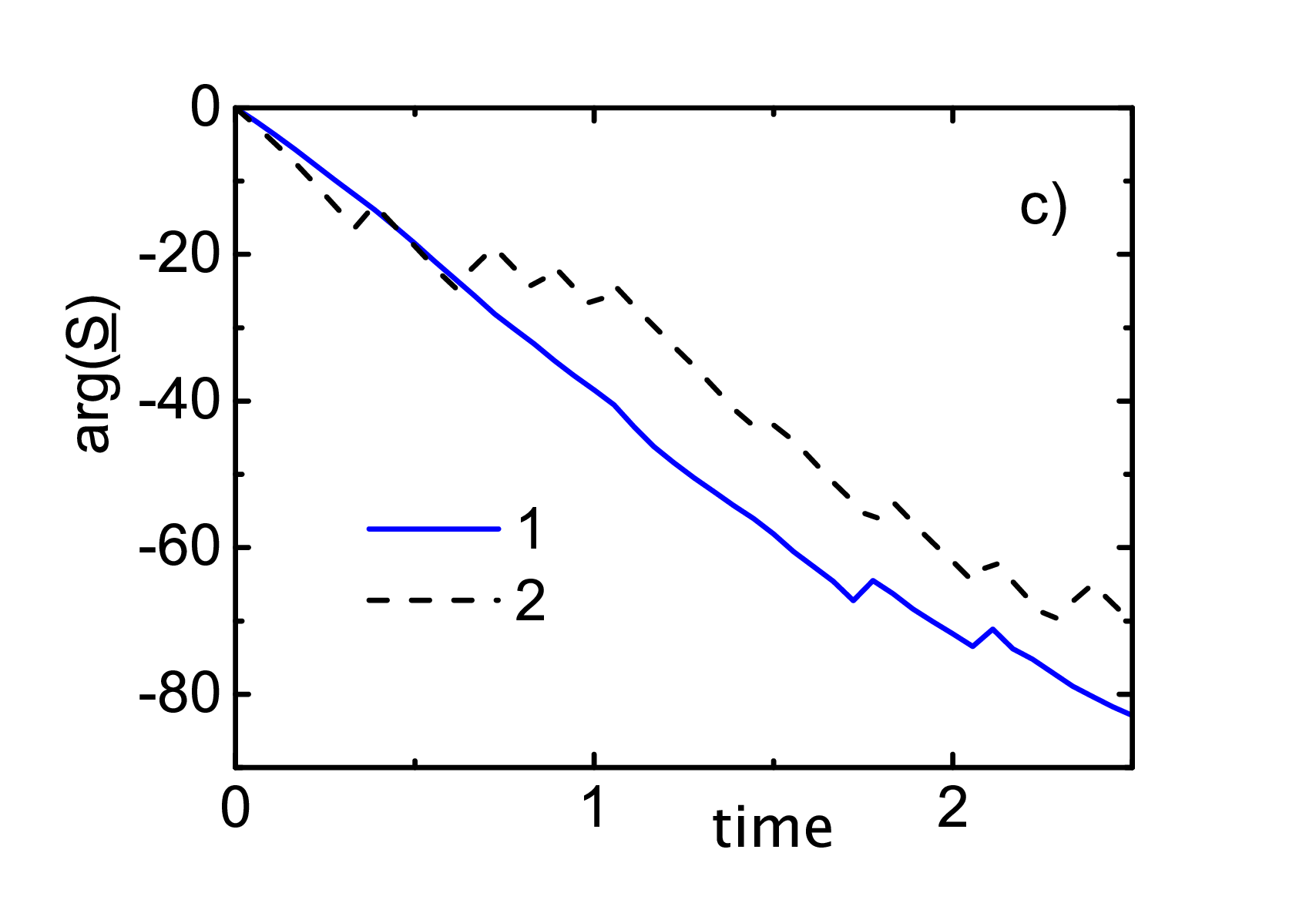}	
	%	%	\vskip 0.3cm
	%	%\hspace{.2cm}
	\caption{(Color online) 
		The $S(|k|, \omega)$ (panel a)) and its inverse Fourier transform $\underline{S}(|k|, t )=|\underline{S}| e^{{\rm i} arg(\underline{S}) }$ 
		(panel b) and panel c)) at $T =0.01 Ry$ $r_s=20$ and  $|k|=0.35$ in conditional units. 
		Lines: 1/2 - uncorrelated/interacting electron pairs. 	
		\label{harm2}	}
\end{figure} 

\begin{figure}[htp] 
	\centering
	%	\vspace{0.3cm}
	\includegraphics[width=0.32\columnwidth,clip=true]{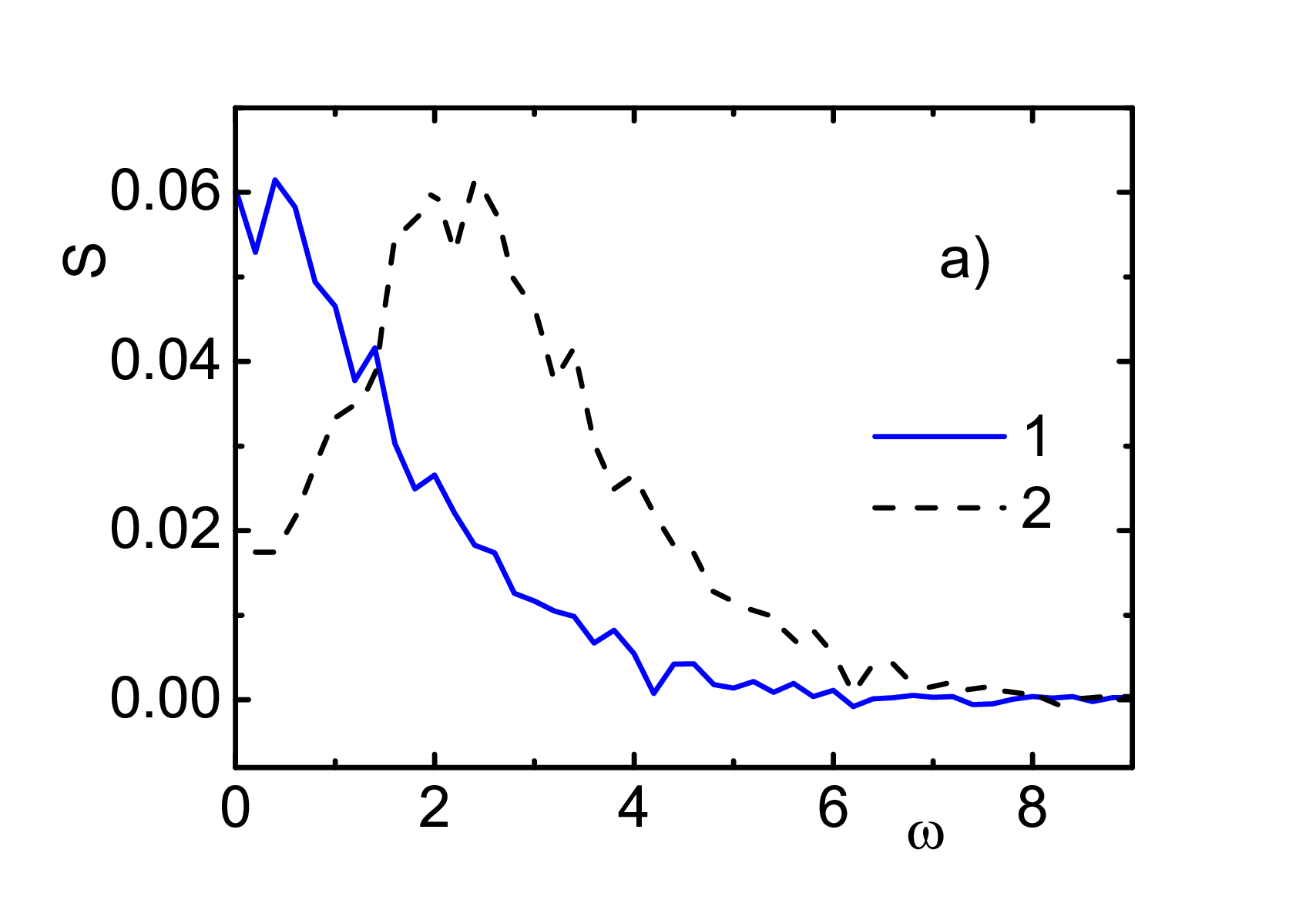}		
	\includegraphics[width=0.32\columnwidth,clip=true]{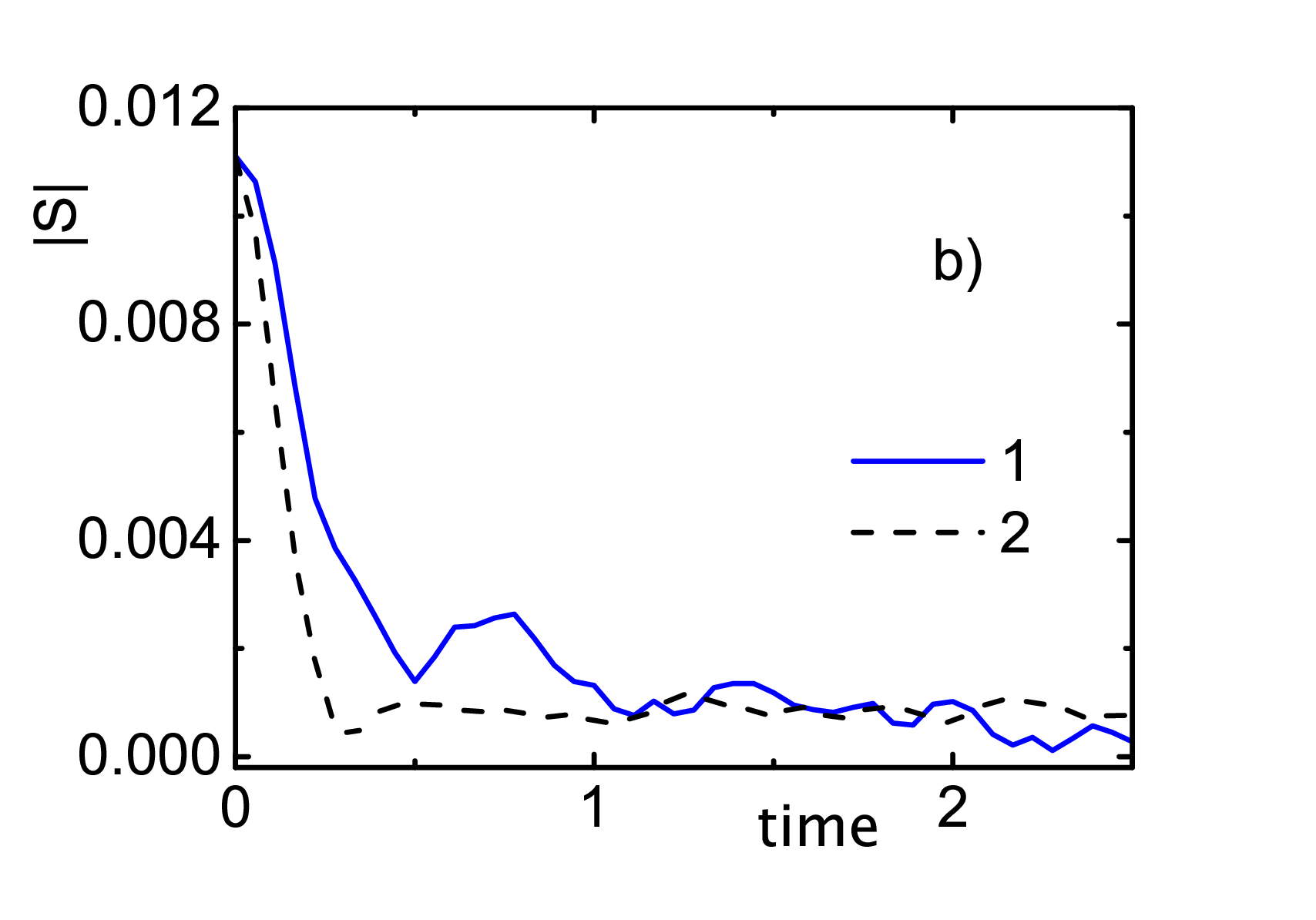}	
	\includegraphics[width=0.32\columnwidth,clip=true]{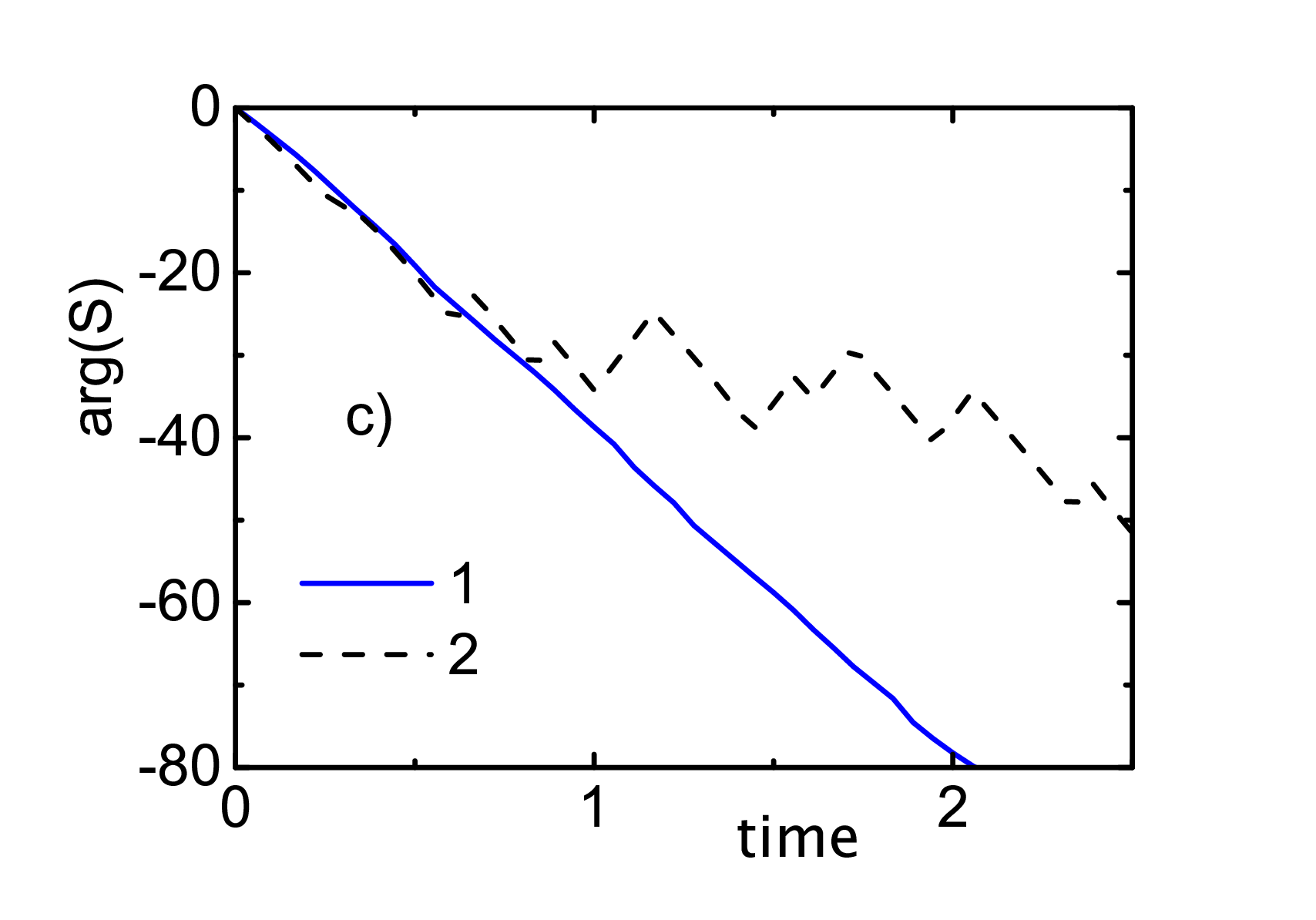}	
	%	%	\vskip 0.3cm
	%	%\hspace{.2cm}
	\caption{(Color online) 
		The $S(|k|, \omega)$ (panel a)) and its inverse Fourier transform 
		$\underline{S}(|k|, t )=|\underline{S}| e^{{\rm i} arg(\underline{S}) }$ 
		(panel b) and c)) at $T =0.07 Ry$ $r_s=10$ and  $|k|=0.35$ in conditional units.  
		Lines: 1/2 - uncorrelated/interacting electron pairs. 
		\label{harm3}
	}
\end{figure}

Figures~\ref{harm2}, \ref{harm3} present ESCF  at higher temperatures  
in comparison with Figures~\ref{harm}  
for two models: line 1--uncorrelated electron pairs accounting only factors 2) and 3),  
line 2--interacting electron pairs accounting factors 1), 2) and 3). 

Comparison of the correlation functions of entangled state $S(|k|, \omega)$ on Fig.~\ref{harm}a) (line 1)  and Fig.\ref{harm2}a) (lines 1 and 2) 
demonstrates a strong transformation of the oscillation spectrum at fixed density but increasing temperature (factor 3) and 
due to interactions described by the factor 1. 

As it follows from comparison of the time entangled state correlation functions 
$\underline{S}(|k|, t )=|\underline{S}| e^{{\rm i} arg(\underline{S}) }$  
(being the inverse Fourier transform of  ESCF)
on Fig.~\ref{harm}b) (line 1)  and Fig.\ref{harm2}b) (lines 1, 2) 
at increasing temperature and fixed density results in decreasing correlation time of entangles states. At the same physical conditions the 
correlation time for uncorrelated electron pairs Fig.~\ref{harm2}b) (line 1) is longer in comparison with one for interacting electron entangled pairs   
Fig.~\ref{harm2}b) (line 2). 

Comparison of Fig.~\ref{harm}c) (line 1)  and Fig.\ref{harm2}c) (lines 1 and 2) shows that  difference in averaged in canonical ensemble 
effective frequency of oscillations of $\underline{S}$  $(|k|, t)$ (time derivative of phase $arg(\underline{S})$) 
for uncorrelated electron pairs and interacting ones is not very impressive but have some peculiarities corresponding to decay of entangled states. 

Figure~\ref{harm3} present the ESCF and TESCF at much higher density and temperatures as above for uncorrelated and interacting electron pairs.  
Now difference in dynamic spectrum of entangled state of uncorrelated and interacting electron pairs is more impressive (Fig.~\ref{harm3}a)(lines 1 and 2)).  
The decoherence time and the effective  frequency  of  oscillations is smaller for interacting electron pairs in comparison with uncorrelated ones.    
(lines 1 and 2 Fig.~\ref{harm3} b) and c).   Effective frequency of oscillations for uncorrelated electron pairs and interacting ones have as above some peculiarities 
corresponding to oscillating decay of entangled states. The averaged in canonical ensemble effective frequency  of oscillations on the whole 
becomes smaller due to the strong increasing  coupling of many entangled pairs. 

\section{Discussion}\label{concl}  
Quantum entanglement expresses the non-locality to quantum mechanics, while 
decoherence is the mechanism by which entanglement is destroyed producing a quantum to classical transition due to the coupling between a quantum system and its environment. 
The decreasing of entanglement may induce failure of the different quantum devices.  
The quantification of entanglement still stands an open issue \cite{hornberger2009introduction}. 

In this work, we present a new path integral \textit{real--time} quantum dynamics approach 
applied to the path integral Monte Carlo simulations of entanglement in quantum many--body systems, 
in particular the quantum dynamics of the pair entangled states and their decay in many--body systems.  

Our approach combines the Heisenberg and Wigner formulation of quantum mechanics, path integral technique 
and generalized  Wiener--Khinchin theorem 
applied to spectral densities of the paths in path integrals. 
The correlation function has been expressed through the Weyl symbols of the density--density operator of the entangled electron pairs 
and the generalization of the Wigner function.  

The developed approach was applied to consider action of thermalization and interparticle 
interaction on the dynamic spectrum and the decoherence  of the pair electron entangled states.   
Presented results include the spin--resolved radial distribution functions, the entangled state correlation function, 
the time entangled state correlation function and estimations of the decoherence time of the entangled pairs 
of electron due to  thermalization and interparticle interaction. 

The developed method offers a wider possibility in treatment dynamic properties as it is formulated in the phase space and 
explicitly uses the momenta and coordinates of particles. 
The developed WPIMC approach will open up prospects for future research and  
can be also useful for many aspects of strongly coupled plasma and extreme states of matter.

\section*{Acknowledgements}

We thank Prof. M.\,Bonitz for valuable discussions. 
The authors acknowledge the JIHT RAS Supercomputer Centre and the Shared Resource Centre ``Far Eastern Computing Resource'' IACP FEB RAS
for providing computing time. 

%\backmatter
%\begin{appendices} 
%\end{appendices}

%%%%%%%%%%%%%%\bibliography{ueg2020.bib} 
%%%%\bibliography{sn-bibliography}% common bib file
%\input sn-article.bbl



\end{document}